\documentclass[preprint2]{aastex}

\shortauthors{Guinan et al.}
\shorttitle{The evolved B-type eclipsing binary V380 Cyg}

\begin{document}

\title{Eclipsing Binaries as Astrophysical Laboratories:\\
Internal Structure, Core Convection and Evolution of the 
B-star Components of V380 Cygni}

\author{Edward F. Guinan\altaffilmark{1}, Ignasi Ribas\altaffilmark{1,2},
Edward L. Fitzpatrick\altaffilmark{1}, \'Alvaro Gim\'enez\altaffilmark{3,4},
Carme Jordi\altaffilmark{2}, George P. McCook\altaffilmark{1}, and Daniel M. 
Popper\altaffilmark{5,6}}

\altaffiltext{1}{Department of Astronomy \& Astrophysics, Villanova University, 
Villanova, PA 19085, USA. E-mail: edward.guinan@villanova.edu,
iribas@ast.villanova.edu, fitz@ast.villanova.edu, george.mccook@villanova.edu}

\altaffiltext{2}{Departament d'Astronomia i Meteorologia, Universitat de 
Barcelona, Av. Diagonal, 647, E-08028 Barcelona, Spain. E-mail: carme@am.ub.es}

\altaffiltext{3}{Laboratorio de Astrof\'{\i}sica Espacial y F\'{\i}sica 
Fundamental, Apartado 50727, E-28080 Madrid, Spain. E-mail: ag@laeff.esa.es}

\altaffiltext{4}{Instituto de Astrof\'{\i}sica de Andaluc\'{\i}a, CSIC, 
Apartado 3004, E-18080 Granada, Spain}

\altaffiltext{5}{Division of Astronomy and Astrophysics, UCLA, CA 90095-1562, 
USA}

\altaffiltext{6}{Deceased}

\begin{abstract}

New photometric solutions have been carried out on the important eccentric
eclipsing system V380 Cygni (B1.5~II-III + B2~V) from $UBV$
differential photoelectric photometry obtained by us. The photometric elements
obtained from the analysis of the light curves have been combined with the 
spectroscopic solution recently published by Popper \& Guinan and have led to 
the physical properties of the system components. The effective temperature of 
the stars has been determined by fitting IUE UV spectrophotometry to Kurucz 
model atmospheres and compared with other determinations from broad-band and 
intermediate-band standard photometry. The values of mass, absolute radius, and 
effective temperature, for the primary and secondary stars are: $11.1\pm
0.5$~M$_{\odot}$, $14.7\pm0.2$~R$_{\odot}$, $21\,350\pm400$~K, and $6.95\pm
0.25$~M$_{\odot}$, $3.74\pm0.07$~R$_{\odot}$, $20\,500\pm500$~K, respectively. 
In addition, a re-determination of the system's apsidal motion rate has been 
done from the analysis of 12 eclipse timings obtained from 1923 to 1995. The
apsidal motion study yields the internal mass distribution of the more 
luminous component. Using stellar structure and evolutionary models with modern 
input physics, tests on the extent of convection in the core of the more 
massive B1.5 II-III star of the system have been carried out. Both the 
analysis of the $\log g - \log T_{\rm eff}$ diagram and the apsidal motion 
study indicate a star with a larger convective core, and thus more centrally 
condensed, than currently assumed. This has been quantified in form of an 
overshooting parameter with a value of $\alpha_{\rm ov}\approx 0.6\pm0.1$. 
Finally, the tidal evolution of the system (synchronization and circularization
times) has also been studied.

\end{abstract}


\section{Introduction}

The bright eclipsing binary V380 Cyg (HR~7567; HD~187879; HIP~97634;
$V_{max}$ = 5.68; $P=12.426$~days; B1.5~II-III + B2~V; Hill \& Batten 1984) 
has properties that make it an important ``astrophysical laboratory'' for 
studying the structure and evolution of massive stars. In particular, the
extent of convection in the stellar core and the internal
mass distribution of the primary component can be probed because the system 
has an eccentric orbit with a well-established apsidal motion rate.
Also, V380~Cyg can provide independent measures of the initial fractional 
helium abundance of the system ($Y$), which is an important and fundamental 
quantity but empirically difficult to measure. Accurate fundamental physical 
properties of the components (e.g., mass, radius, effective temperature, 
etc.) are however required to carry out such analyses.

V380~Cyg consists of an evolved, more massive, and more luminous primary 
component and a main sequence secondary star. It has an eccentric orbit
($e=0.23$) and an orbital period of 12.426~days. Several spectroscopic
studies have been carried out to determine its orbital properties and
the masses, temperatures, and luminosities of the component stars (see Batten 
1962; Popper 1981; Hill \& Batten 1984; Lyubimkov et al. 1996). The first
light curve of the system by Kron (1935), as well as the more modern
photometry by Semeniuk (1968) and Battistini, Bonifazi, \& Guarnieri (1974), 
revealed shallow eclipses with depths of $\sim0.12$~mag and $\sim0.09$~mag for 
primary and secondary minima, respectively. Moreover, these photometric
measurements show that the secondary eclipse is displaced from
$0\fp5$, indicating an eccentric orbit, while changes in the
displacement between primary and secondary minima indicate the
presence of apsidal motion with a period of about 1500 yrs. 

As discussed by Gim\'enez (1984) and Gim\'enez, Claret, \& Guinan (1994), 
V380~Cyg is an ideal binary system for the study of convective overshooting 
in the cores of massive stars because of the evolutionary stages of its 
component stars. Because of the eccentric orbit and the eclipsing nature of
the binary, it is possible to determine the apsidal motion rate. From this,
additional constraints on the internal mass distribution and the evolutionary
state of an evolved massive star also can be established.

Because V380~Cyg provides potentially important tests of stellar structure and
evolution, we have carried out new high signal-to-noise spectroscopic and 
photometric observations of the system and 
performed a new, detailed investigation of its properties. In \S \ref{sec:new} 
we present the new observations. The effective temperature determination, based
on both standard photometry and UV/optical spectrophotometry, is discussed in 
\S \ref{sec:teff}. In \S \ref{sec:modLC} we concentrate on the analysis of 
the light curves. \S \ref{sec:aps} is devoted to the study of eclipse 
timings, leading to an accurate determination of the apsidal motion rate. The 
properties of the system components are compared with the predictions of 
theoretical models in \S \ref{sec:mod}. The tidal evolution of the system 
(circularization and synchronization times) is analyzed in \S \ref{sec:tidal}. 
Finally, the main conclusions of our study are presented in \S \ref{sec:conc}.

\section{Observational Data} \label{sec:new}

New high-quality spectroscopic and photometric observations have been 
collected for V380~Cyg. The following sections contain a brief description 
of these data.

\subsection{Spectroscopy}

V380~Cyg is difficult to resolve spectroscopically into a double-line 
binary because of the large luminosity difference between the component stars. 
Therefore, very high signal-to-noise and high-resolution spectroscopic 
observations are needed to measure the radial velocity shifts of the faint 
lines of the secondary star. Such high signal-to-noise and high-resolution 
spectroscopy was recently secured by us and led to a new, accurate radial 
velocity curve, which was carefully analyzed in a previous paper (Popper 
\& Guinan 1998; hereafter Paper~I).  The spectroscopic solution obtained 
in Paper I (basically velocity semi-amplitudes and systemic velocity) has 
been adopted in the present study.

\subsection{Photometry}            

$UBV$ differential photoelectric photometry of V380~Cyg was carried out
from 1988 to 1997 using robotic, automatic photoelectric telescopes
(APTs) located on Mt. Hopkins, Arizona. From 1988 to 1989 the
photometry was conducted with the Phoenix-10 APT telescope, a 25-cm
robotic telescope equipped with a 1-P21 photometer and filters matched
to the standard $UBV$ system. During 1995--1997 the photometry was
conducted using the Four College Consortium (FCC) 0.8-m APT. The FCC
APT is equipped with filters closely matching the standard $UBV$ system
and a refrigerated Hamamatsu photoelectric detector. Differential photometry 
was obtained in which HD~189178 (B5~V, $V=5.47$) was employed as the
comparison star, while HD~188892 (B5~IV, $V=4.94$) served as the check
star. No evidence of significant light variations (down to the few
milimagnitude level) was found for the comparison--check star sets. 
The observations were reduced using photometric reduction programs at 
Villanova University. Differential extinction corrections were applied, 
but these corrections were typically very small since the comparison star 
is within 2~deg. of V380~Cyg and the stars were never observed at high air 
mass. The photometric measurements are listed in Table \ref{tab:phot}.

\placetable{tab:phot}

\begin{figure*}
\epsscale{1.40}
\plotone{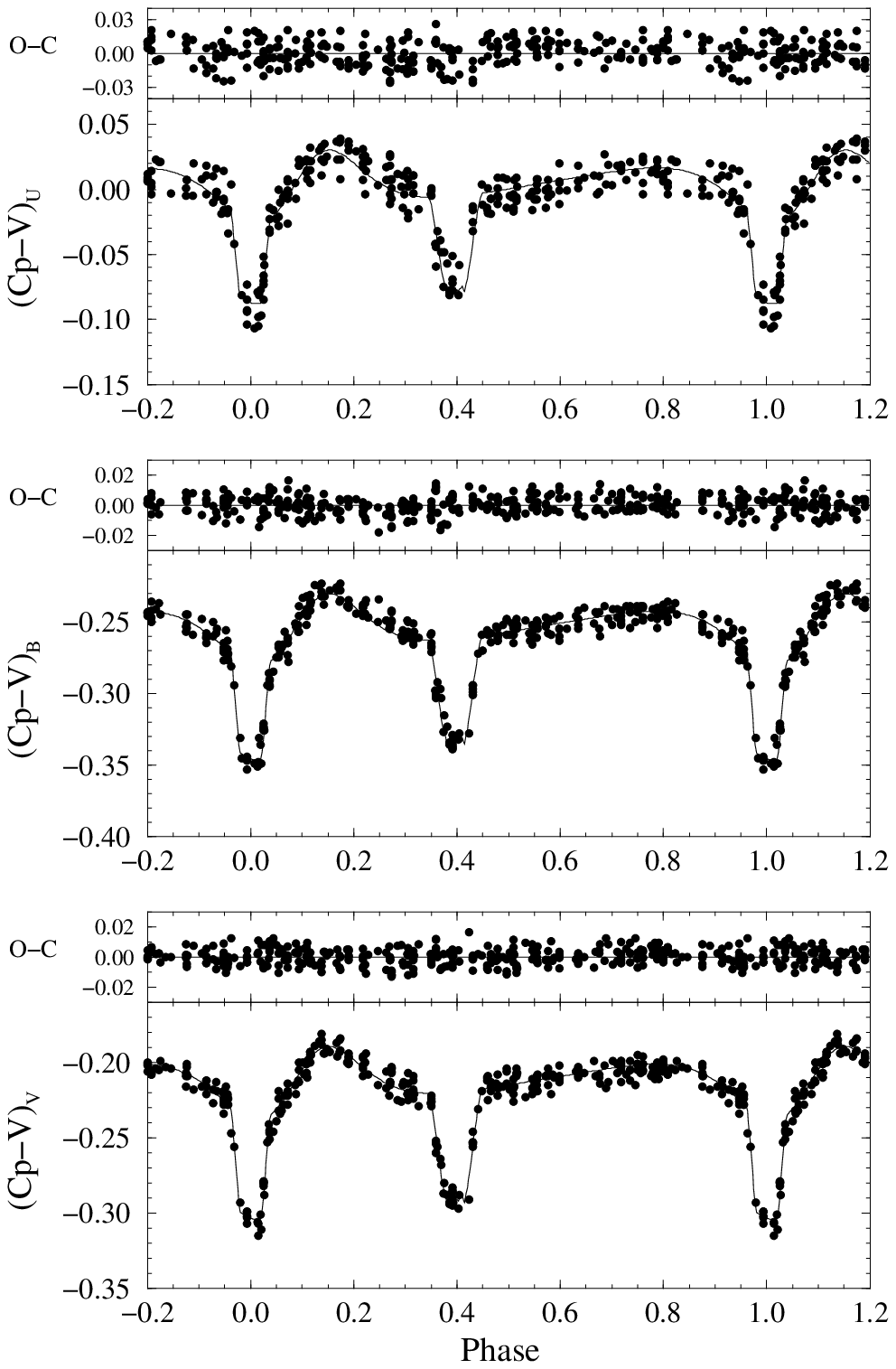}
\figcaption[fig1.eps]{Light curve fit to the observed $U$, $B$ and $V$
Comparison$-$Variable (Cp$-$V) differential photometry of V380~Cyg. Also
shown are the Observed$-$Computed (O$-$C) residuals. The orbital phase
was computed according to the following ephemeris: $T_{\rm Min I} =
{\rm HJD}2441256.544 + 12.425719\,E$. \label{fig:lc}}
\end{figure*}

$U$, $B$, and $V$ differential light curves are shown in Figure \ref{fig:lc}.
We adopted the following ephemeris for computing the orbital phase:
$T_{\rm Min I} = \mbox{HJD}2441256.544 + 12.425719\,E$ (see \S \ref{sec:aps}).
As can be seen in the plots, the shallow secondary minimum 
occurs at $0\fp403$. Also, the light curves show out-of-eclipse brightness 
variations with maximum light occurring near $0\fp15$. The displacement of the 
secondary eclipse from phase $0.5$, as noted in \S 1 above, is caused by
the eccentric orbit of the system, while the outside eclipse light variations 
arise primarily from the changing tidal distortion, chiefly from the larger 
star as it moves in the eccentric orbit. According to the recent spectroscopic 
solution of Paper~I, periastron passage occurs near $0\fp15$ where the stars 
reach their greatest tidal deformations.

\placefigure{fig:lc}

\section{Temperature Determination} \label{sec:teff}

A complete analysis of the V380~Cyg system requires precise measurements of
the stellar effective temperatures ($T_{\rm eff}$).  For this purpose, we
explored the use of published T$_{\rm eff}$ vs. intrinsic color
calibrations --- utilizing several different photometric systems ---
and also we performed a detailed study of the shape of the UV/optical stellar
energy distribution.  Both these methods for determining effective temperatures
are discussed below.

Intermediate-band Str\"omgren and Vilnius photometric indices for
V380~Cyg are available from the literature and are listed in Table
\ref{tab:teff} along with our new measurements in the Johnson $UBV$
system. The Johnson photometry was secured out-of-eclipse, when both
stars were fully visible. The phases at which the Str\"omgren and
Vilnius measurements were made is not known, although this is not
likely to be a critical problem for V380~Cyg because the secondary
component is much less luminous than the primary component and the
temperatures and color indices of the two stars are similar.
A summary of the reddening and $T_{\rm eff}$ results for the three
photometric systems is given in Table \ref{tab:teff}.  An {\em a
priori} estimate of the expected uncertainties in these results is
difficult, but clearly depends on both random errors in the stellar
photometry and systematic errors in the various calibrations
themselves. It is clear, however, that the level of agreement among
the various systems is unsatisfactory, with a range of over 4000 K and
a standard deviation of $\sim$1800 K --- nearly 10\% of the value of
$T_{\rm eff}$ itself.

\placetable{tab:teff}

In view of the discrepancies among the photometric results for
$T_{\rm eff}$, we employed a different technique for determining this
important quantity. We modeled the observed shape of the UV/optical
energy distribution of V380~Cyg using theoretical spectra generated
with the Kurucz's {\em ATLAS9} model atmosphere program. This
analysis was developed by Fitzpatrick \& Massa (1999; hereafter FM99)
and has been applied previously to an eclipsing binary system by Guinan
et al. (1998).

For a binary system, the observed energy distribution depends on the
surface fluxes of the binary's components and on the attenuating
effects of distance and interstellar extinction. This relationship can
be expressed as:

{\tiny 
\begin{equation} \label{basic1}
f_{\lambda\oplus} =\left(\frac{R_{\rm P}}{d} \right)^2 \left[F_{\lambda}^{\rm P} + \left(\frac{R_{\rm S}}{R_{\rm P}}\right)^2 F_{\lambda}^{\rm S}\right] \times 10^{-0.4 E(B-V) [k(\lambda-V) + R(V)]}
\end{equation}}
where $F_{\lambda}^{\rm P}$ and $F_{\lambda}^{\rm S}$ are the surface fluxes 
of the primary (P) and secondary (S) components, $R_{\rm P}$ and $R_{\rm S}$
are the absolute radii, and $d$ is the distance to the binary. The last term 
carries the extinction information, including  $E(B-V)$, the normalized 
extinction curve $k(\lambda-V)\equiv E(\lambda-V)/E(B-V)$, and the ratio of
selective-to-total extinction in the $V$ band $R(V) \equiv A(V)/E(B-V)$.

The modeling process consists of a non-linear least squares
determination of the optimal values of all the parameters which
contribute to the right side of equation \ref{basic1}. In principle,
the problem can involve solving for two sets of Kurucz model parameters
(i.e., one set of $T_{\rm eff}$, surface gravity, metallicity $[m/H]$, and
microturbulence velocity $v_{\rm micro}$ values for each star), the ratios
$(R_{\rm P}/d)^2$ and $R_{\rm S}/R_{\rm P}$, $E(B-V)$, $k(\lambda-V)$, and 
$R(V)$. For V380 Cyg, however, a number of simplifications can be made: (1) 
the temperature ratio of the two stars is known from the light curve
analysis (see \S \ref{sec:modLC} below); (2) the surface gravities are known 
from the combined result of the light and radial velocity curve analyses (see 
\S \ref{sec:modLC}); (3) $[m/H]$ can be assumed to be identical for both 
components; (4) $v_{\rm micro}$ can be assumed to be 0 for the much fainter 
secondary star; (5) the ratio $R_{\rm S}/R_{\rm P}$ is known from the light 
curve analysis (see \S \ref{sec:modLC}); (6) the properties of 
normalized UV/optical extinction are constrained by a small number of 
parameters (Fitzpatrick \& Massa 1990; Fitzpatrick 1999); and (7) the 
standard value of $R(V) = 3.1$ may be assumed, without affecting the 
$T_{\rm eff}$ determination.

\begin{figure*}
\epsscale{2.00}
\plotone{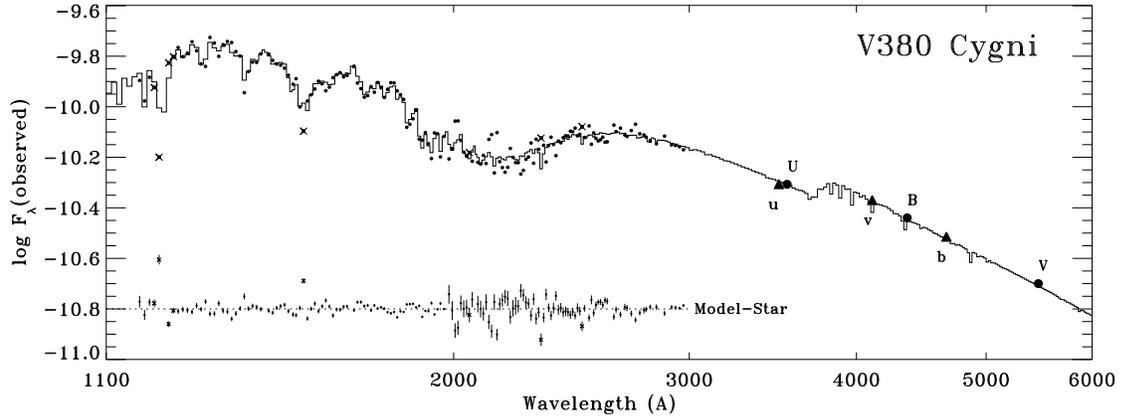}
\figcaption[fig2.eps]{Kurucz atmosphere model fit (solid line) to IUE
SWP+LWP spectrum (dots) of V380~Cyg. The cross symbols indicate wavelength
regions that were not used in the fits chiefly because of the presence of
contaminating interstellar medium features. The filled circles and triangles
are Johnson and Str\"omgren fluxes respectively. The residuals of the fit
are shown in the lower part of the figure. \label{fig:iue}}
\end{figure*}

Adopting these simplifications, we modeled the observed UV/optical
energy distribution of V380~Cyg by solving for the best-fitting values
of $T_{\rm P}$, $[m/H]_{\rm PS}$, $v_{\rm micro}^{\rm P}$, $(R_{\rm P}/d)^2$, 
$E(B-V)$, and $k(\lambda-V)$. The observational data used in the analysis 
includes the Johnson and Str\"omgren optical photometry listed in Table
\ref{tab:teff} and UV spectrophotometry from the {\it International
Ultraviolet Explorer (IUE)} satellite.  The {\em IUE} data consist of a
low-dispersion short-wavelength spectrum (SWP38553; 1150~\AA~to
1980~\AA) and a low-dispersion long-wavelength spectrum (LWP17712;
1980~\AA~to 3000~\AA), both secured out-of-eclipse.  These data were
obtained from the {\it IUE} Final Archive and thus were processed using
the {\it NEWSIPS} system (Nichols \& Linksy 1996).  We utilized the
algorithms of Massa \& Fitzpatrick (2000) to correct these {\it
NEWSIPS} data for strong thermal and temporal systematics and to place
them on the {\it Hubble Space Telescope} absolute UV/optical
calibration system (Bohlin 1996).  The final {\it IUE} spectrum was
then re-sampled to match the wavelength binning of the {\em ATLAS9}
models (see FM99). This binned spectrum can be seen in Figure \ref{fig:iue} 
(small filled circles).  The presence of the 2175 \AA\/ interstellar medium 
extinction feature is clear, as are the strong stellar absorption
features in the 1400 \AA, 1500 \AA, and 1900 \AA\ regions characteristic
of evolved early B-type stars. For display purposes, we also show the
Johnson and Str\"omgren magnitudes, converted to fluxes, in Figure
\ref{fig:iue} (large filled circles and triangles).  However, in the
fitting procedure the photometric indices themselves (i.e., $V$, $B-V$,
$U-B$, $b-y$, $m_1$, $c_1$, $\beta$) were compared with synthetic values for 
the models. The calibration of the synthetic photometry is discussed by
FM99 and in a forthcoming paper by E. L. Fitzpatrick \& D. Massa (in 
preparation).
 
The best-fitting, reddened, and distance-at\-tenuated {\em ATLAS9} model
of V380~Cyg's energy distribution is shown in Figure \ref{fig:iue}
(histogram-style curve) overplotted on the V380~Cyg observations.  In
the UV region, the residuals to the fit (``Model -- Star'') are shown
below the energy distribution, with the vertical bars indicating the
statistical error in the {\em IUE} data. The major discrepancies occur
near the interstellar Ly $\alpha$ line (1215 \AA) and the C IV stellar
wind line (1550 \AA). These regions were excluded from the fit, as
indicated by the crosses over the data points. Several other data
points, corresponding to spurious feature in the models were also
excluded (see FM99). (Note that the fit appears poorest in the 2000 to
2300 \AA\/ region because of the low signal-to-noise level of the
data.)

\placefigure{fig:iue}

The parameters characterizing the fit are listed in Table \ref{tab:prop}. The
values of $T_{\rm eff}$ and $E(B-V)$ are within the range found from the
photometric analysis (see Table \ref{tab:teff}) but are much more precisely
determined. (The synthetic photometry of the best-fitting model
reproduces all the Johnson and Str\"omgren photometric indices to
within the expected errors of the data.) The attenuation factor of the energy 
distribution is also derived from the analysis, and it yields the distance 
to the system (Table \ref{tab:com}) when combined with the absolute radius 
of the primary component (see Guinan et al. 1998 for further discussion on 
this issue). The values of $[m/H]$ and $v_{\rm micro}$ bear some comment. 
FM99 found that in B-type stars with significant stellar winds, correlated 
increases in the best-fitting values of $v_{\rm micro}$ and decreases in 
$[m/H]$ occur. These combinations reproduce the observed opacity features 
very well, but it is likely that neither parameter faithfully represents a 
corresponding stellar property.  This probably arises because the stellar 
opacity features form in a region with a weak velocity gradient, while the 
models explicitly assume an atmosphere in hydrostatic equilibrium. (See FM99
for additional discussion.) The implication for this work is that the
value of $[m/H]$ derived from this analysis is almost certainly an
underestimate of the true metallicity of V380~Cyg. For further analysis,
we adopt a metal abundance of $[m/H]=-0.2$ based on recent analyses of 
nearby B-type stars (Kilian, Montenbruck, \& Nissen 1994), solar-type stars 
(Edvardsson et al. 1993) and interstellar medium studies (Meyer, Jura, \& 
Cardelli 1998). 

Note that this fitting procedure relied on results generated by the
light and radial velocity curve analyses described below, while these
other analyses rely (albeit very weakly) on the $T_{\rm eff}$'s determined
here. We thus performed these three analyses in an iterative fashion
until a self-consistent set of parameters resulted.
 
\section{Modeling the Light Curves} \label{sec:modLC}

The light curves were solved using an improved version of the Wilson \& 
Devinney (1971; hereafter WD) program (Milone, Stagg, \& Kurucz 1992; Milone
et al. 1994) that includes Kurucz {\em ATLAS9} atmosphere models for the 
computation of the stellar radiative parameters. A detached configuration (as 
suggested by the eccentric orbit) with coupling between luminosity and 
temperature was chosen when running the solutions. Both reflection and 
proximity effects (i.e., tidal distortion) were taken into account, 
because the light curves clearly show their significance. The
bolometric albedo and the gravity brightening coefficients were set to a value
of 1.0 as usually adopted for the radiative envelopes of these early-type
stars. The mass ratio ($q\equiv M_{\rm S}/M_{\rm P}$) was fixed to the 
spectroscopic value of $q=0.626$ and the temperature of the primary star was 
set to 21\,350~K. Finally, pseudo-synchronization (synchronization at 
periastron) was assumed for both components (see Sect. \ref{sec:tidal}).

All three $UBV$ light curves were solved simultaneously to achieve a single, 
mutually consistent, solution. In this simultaneous solution, the relative 
weight of each curve, which has significant importance, was set according to 
the r.m.s. residual of the comparison$-$check star sets (i.e., 0.015, 0.006, 
and 0.006 for $UBV$, respectively). The free parameters in the light curve 
fitting were: the eccentricity ($e$), the longitude of the periastron 
($\omega$), the phase offset ($\phi_{\circ}$), the orbital inclination
($i$), the temperature of the secondary ($T_{\rm S}$), the gravitational
potentials ($\Omega_{\rm P}$ and $\Omega_{\rm S}$), and the luminosity of
the primary ($L_{\rm P}$). Depending on the initial conditions, two different
sets of parameters providing good fits to the observations were found.
There is not sufficient information in the light or radial velocity curves to
discriminate between these two possible solutions, and additional data sources
have to be considered to break the degeneracy. Since the luminosity ratio
between the components predicted by the two solutions is very different
($L_{\rm S}/L_{\rm P}\approx0.05$ and $L_{\rm S}/L_{\rm P}\approx0.5$), the
spectroscopic light ratio ($L_{\rm S}/L_{\rm P}\approx 0.05$, from Paper~I)
was used to confirm the first solution as the physically preferable one.

An automatic procedure based on {\sc midas} scripts and {\sc fortran} programs 
was employed to execute the WD code. At each iteration, the differential 
corrections were applied to the input parameters to build the new set of 
parameters for the next iteration. The result was carefully checked in order 
to avoid possible unphysical situations (e.g., Roche lobe filling in detached 
configuration). A solution was defined as the set of parameters for 
which the differential corrections suggested by the WD code were smaller than 
the standard errors during three consecutive iterations. However, when a 
solution was found, the program did not stop. Instead, it was kept running for 
a given number of iterations, 100 in our case. This was done to test 
the stability of the solutions, to evaluate their scatter, and to check for 
possible spurious solutions. For V380~Cyg, the program rapidly converged to 
a very stable minimum and 47 individual solutions resulted from the 100 
iterations. A summary of the best-fitting parameters to the light curve
is presented in Table \ref{tab:prop}.

We also considered -- and ultimately rejected -- the possibility of a third 
light contribution to the system. When 
searching both the SIMBAD database and the Tycho catalogue (ESA 1997) no 
objects were found within a radius of 1~arcminute centered at the position 
of V380~Cyg. The system was observed by the Hipparcos mission and there is 
no indication of visual duplicity down to the spatial and intensity resolution 
of the satellite ($\rho \approx 0.1-0.15$~mas and $\Delta {\rm mag}
\approx3.5-4$). Moreover, no evidence of stationary spectral lines belonging 
to a third component are observed in the high $S/N$ spectra used for the 
spectroscopic study. Since the contribution of the secondary component (whose 
lines are seen in the spectra) to the total flux of the system in the 
considered wavelengths is only about 6 percent, we therefore conclude that any 
third light can be safely neglected. 

Additional WD runs were performed starting from the same initial conditions
but fixing $T_{\rm P}$ to its proper value $\pm$1$\sigma$. This showed that
our results from the light curve solution are insensitive to temperature 
variations within this range.

\placetable{tab:prop}

A remark should be made at this point concerning the physical meaning of 
the primary and the secondary eclipses. Since the massive and larger 
component is also somewhat hotter (and so it has slightly higher surface 
flux), one would therefore expect the primary (and deeper) eclipse to be a 
transit of the smaller star in front of the bigger one. However, this is 
not the case for V380~Cyg, and the primary eclipse turns out to be an 
occultation of the smaller and cooler component. Such a behavior can be 
explained in terms of the eccentric orbit and the variations induced by the 
proximity effects, which can be seen clearly as an out-of-eclipse wave-like 
variation in Figure \ref{fig:lc}. Indeed, eccentricity-induced proximity 
effects produce minimum flux near the phase of the transit, making it the 
primary eclipse.

\begin{figure*}
\epsscale{1.50}
\plotone{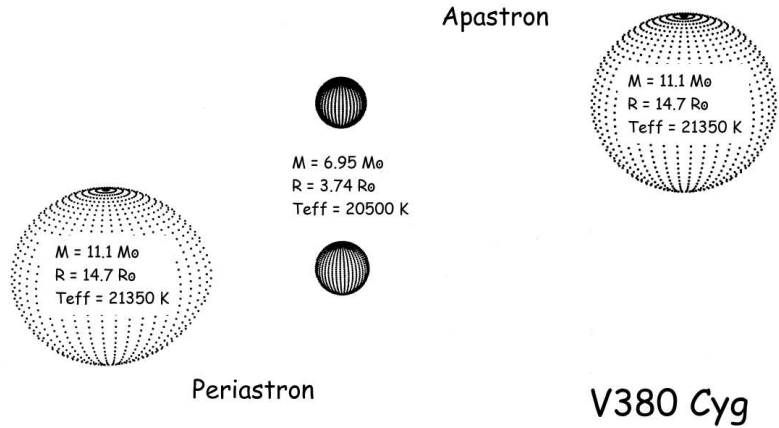}
\figcaption[fig3.eps]{3-D picture of V380~Cyg at periastron and apastron.
Notice the slight deformation of the primary component during periastron.
\label{fig:3D}}
\end{figure*}

The uncertainties in the parameters presented in Table \ref{tab:prop} were
carefully evaluated by means of two different approaches. On the one hand, the
WD program provides the standard error associated with each adjusted parameter.
On the other hand, the error can also be estimated as the r.m.s. scatter of
{\em all} the parameter sets corresponding to the iterations between the first
solution and the last iteration (about 100). This is, 
in general, larger than the r.m.s. scatter of the solutions alone, i.e., those 
that fulfill the criterion previously described. We conservatively adopted 
the uncertainty as twice whichever of the two error estimates was the largest.
The r.m.s. scatters of the residuals in the light curve fit are 0.011 
($n=264$), 0.006 ($n=303$) and 0.005~mag ($n=303$) for $U$, $B$, and
$V$, respectively. Figure \ref{fig:lc} shows the light curve fits to
the observed $U$, $B$, and $V$ differential photometry and the corresponding 
residuals, where no systematic trends are observed. A 3D diagram of V380~Cyg
(drawn to scale) is shown in Figure \ref{fig:3D}, where the tidal deformation 
of the larger primary component near periastron becomes evident. As shown, 
the components are nearly spherical in shape at apastron.

\placefigure{fig:3D}

The results of the light curve fitting were combined with the velocity
semi-amplitudes from the spectroscopic analysis (Paper~I) and the 
UV/optical spectrophotometry study to obtain the absolute dimensions and 
radiative parameters listed in Table \ref{tab:com}. The most recently 
published complete light curve study dates back to Hill \& Batten (1984). 
These authors used the {\sc light}
program to analyze four light curves from various authors and with different
qualities. In general, they determined a fractional radius for the primary
component that is significantly larger than our value, and for the secondary
component this situation is reversed but the difference is much smaller. The
reason for such a discrepancy probably originates in the model used in their 
light curve solution, since the tidal distortion of the primary component is 
strong and also varies due to the eccentric orbit. In this sense, WD
relies on a much more physically rigorous model than {\sc light} and thus, our 
parameters ought to be preferred. Moreover, the fractional radius of the primary
component is very well constrained from the shape of the out-of-eclipse curve
near periastron (0\fp15), a region which is not well covered in the light
curves analyzed by Hill \& Batten (1984). Finally, these authors adopt an
effective temperature for the primary component ($24\,500$~K), based on 
Johnson photometry, which is clearly inconsistent with our detailed analysis. 
However, their temperature ratio is in good agreement with our result.

\placetable{tab:com}

An inspection of the masses and radii obtained reveals a binary system composed
of two massive stars with rather different physical properties: the more 
massive primary has evolved beyond
the main sequence whereas the lower-mass secondary star is still near to the
zero-age main sequence (ZAMS). This situation makes V380~Cyg an extremely
important astrophysical tool since it allows the stellar internal structure
to be probed during a fast-evolving stage. Our determinations of masses and 
radii have accuracies of about 4 percent and 2 percent, respectively. This 
is not as good Andersen (1991) suggests is required for a critical evaluation 
of stellar evolutionary models. However, because of the short evolutionary 
timescales of the primary component of V380~Cyg and the strong dependence on 
the convection parameters, this level of accuracy does allow for important 
tests of the models, as we will show in \S \ref{sec:mod}.

\section{Times of Minima and Apsidal Motion Study} \label{sec:aps}

V380~Cyg is a challenging system from the viewpoint of times of minima 
determinations: both eclipses have long durations (more than 24 hours), they
have nearly-flat bottoms, and their shape is perturbed by the wave-like 
variation seen in the light curves. Therefore, the calculation of the 
exact time of central-eclipse is not straightforward. Because complete 
coverage of one eclipse event is not possible during a single night,
we decided to combine those observations corresponding to a similar
epoch into a light curve and derive the times of the eclipses from
comparison with the best available synthetic model. In our particular 
case, the most natural way of performing such operation is by
considering the APT-Phoenix10 and the FCC 0.8-m APT data separately. The
periods span about 1 and 2.5 years and have around 100 and 200 measurements
per filter, respectively. An additional epoch was retrieved from observations 
carried out during 1977/78 by Dorren \& Guinan (unpublished) at Pahlavi 
Observatory (Shiraz, Iran). The exact time of minimum determination was then 
done by fitting synthetic light curves computed with all the parameters listed 
in Table \ref{tab:prop} to the actual observations at three different epochs.
The resulting eclipse times are presented in Table \ref{tab:ecl}. 
In addition, Table \ref{tab:ecl} lists the eclipse timings from Kron 
(1935), Semeniuk (1968), and Battistini et al. (1974). The mean HJD of each 
data set was adopted as the reference epoch. The secondary eclipse published 
by Semeniuk (1968) was given a much lower weight in the calculations because 
of the poor coverage of the eclipses and thus the high uncertainty in the 
time determination.

\placetable{tab:ecl}

\begin{figure*}
\epsscale{1.20}
\plotone{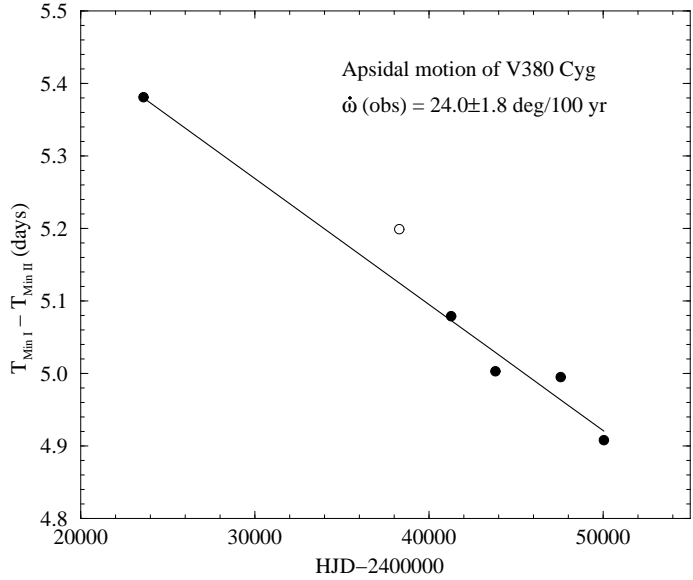}
\figcaption[fig4.eps]{Apsidal motion calculation through a linear fit
to observed time difference between primary and secondary minima. The
slope of the straight line is $1.74\times10^{-5}$. The individual points
have been computed from the times of minima listed in Table \ref{tab:ecl}.
Semeniuk's (1968) value is plotted as an open circle and was not used in
the fit (see text). \label{fig:apsid}}
\end{figure*}

The eclipse timings listed in Table \ref{tab:ecl} allow the determination of 
linear ephemerides independently for primary and secondary eclipses with the 
following results (numbers between parentheses indicate errors in the last 
digits and $E$ is the number of cycles):

{\small
\begin{eqnarray*}
T_{\rm Min I}& =& {\rm HJD}2441256.544(6) + 12.425719(14)\, E \\
T_{\rm Min II}& =& {\rm HJD}2441261.625(5) + 12.425501(7)\, E
\end{eqnarray*}}
The difference in the apparent periods is of course a clear indication of
the presence of a significant apsidal motion. The complete determination of 
the apsidal motion rate of V380~Cyg was accomplished in two steps. First, we 
computed the corresponding time 
differences $T_{\rm Min II} - T_{\rm Min I}$ for each epoch and a linear
least-squares fit to the data clearly showed the presence of a temporal 
decrement of $1.74\times10^{-5}$. The observations of Semeniuk (1968) were not 
used because of the previously mentioned lower accuracy. The result of the 
linear fit is shown in Figure \ref{fig:apsid}, where the value of Semeniuk's 
measurement is plotted with an open circle. The measured slope corresponds 
to an apsidal motion rate of $0.0084$ degrees per cycle, or a period of 
$U=1460$ years, when an orbital eccentricity of $0.234$, an inclination of 
$82.4$ degrees, and a reference longitude of periastron of $\omega_{\circ}=128$ 
degrees (representative of the average epoch) are adopted. In computing the
apsidal motion rate, terms no higher than $e^2$ were used in the relevant 
equations. The slope determination appears to depend critically on Kron's 
(1935) early data point. However, when this measure is excluded, the resulting 
slope is still very similar ($1.68\times10^{-5}$) and within the expected 
uncertainties.

As a second step, because of the relatively large orbital eccentricity,
the method described by Gim\'enez \& Garc\'{\i}a-Pelayo (1983), with equations 
revised by Gim\'enez \& Bastero (1995), was used for a more accurate 
calculation of the apsidal motion rate. When adopting the orbital 
inclination ($82.4\pm0.2$~deg) and the eccentricity ($0.234\pm0.008$) as 
derived from the light curve analysis (see Table \ref{tab:prop}), we 
obtained an apsidal motion rate of $0.0083\pm0.0006$ degrees per cycle, 
equivalent to $\dot{\omega}_{\rm obs}=24.0\pm1.8$ degrees per 100 years or 
an apsidal motion period of $U=1490\pm120$ years. These values are in 
excellent agreement with previous determinations and the position of the 
periastron predicted with these elements for the epoch of the radial velocity 
curve of Paper~I, HJD 2449500, is $\omega=132.4\pm1.5$ degrees also in 
excellent agreement with that analysis ($\omega=133.2\pm3.2$).

\section{Evolutionary Model Study} \label{sec:mod}

Stellar evolutionary theory provides a description of the interior structure
and the observable properties of a star of given initial mass and chemical
composition as a function of age. Significant progress has been made in
recent years in understanding the physical processes that govern the structure
and evolution of stars, including the computation of accurate nuclear reaction 
rates, neutrino emission rates, and atmospheric opacities.

Modeling of the convective energy transport mechanism is probably one of
the major deficiencies of the current stellar structure and evolution
models (see Gim\'enez, Guinan, \& Montesinos 1999 and references therein). 
Because of the 
extreme mathematical complexity of an accurate treatment, convection is still 
described through a very simplified and phenomenological theory: the mixing 
length theory (MLT). When comparing the predictions of the models for stars 
with convective nuclei, an extra amount of convection (overshooting) is needed 
to reproduce the observed properties. The overshooting parameter ($\alpha_{\rm 
ov}$), i.e., the size in pressure-scale height of the over-extension of the 
convective nucleus, cannot be theoretically computed and thus needs to be 
determined from observations. The significance of overshooting is probably 
one of the largest uncertainties in MLT, as well as the possible dependence 
of this parameter on mass, evolution or chemical composition. Some attempts 
(Canuto \& Mazzitelli 1991, 1992; Canuto 1999) have been directed towards the 
development of an essentially parameter-free convection model, but the MLT is 
still being widely used with fairly good results in most applications. 

The eclipsing binaries are powerful tools for testing stellar structure and 
evolution models, since the fundamental properties of the components (masses,
radii, luminosities, etc.) can be accurately determined from observations. 
Several studies testing the significance of the convective overshooting in 
the stellar nucleus using eclipsing binary data (Andersen, Nordstr\"om, \& 
Clausen 1990; Pols et al. 1997; Ribas et al. 1999) have been published. 
Although these studies clearly show an improvement when overshooting is 
included, more quantitative conclusions are difficult to access. This is 
primarily because the effects of overshooting are only observationally relevant 
for evolved stages, especially beyond the TAMS. However, no evolved early-type 
detached eclipsing binaries with accurate determination of absolute dimensions 
were available to date because of observational selection effects. V380~Cyg 
is in this sense an important and unique system. 

\begin{figure*}
\epsscale{1.40}
\plotone{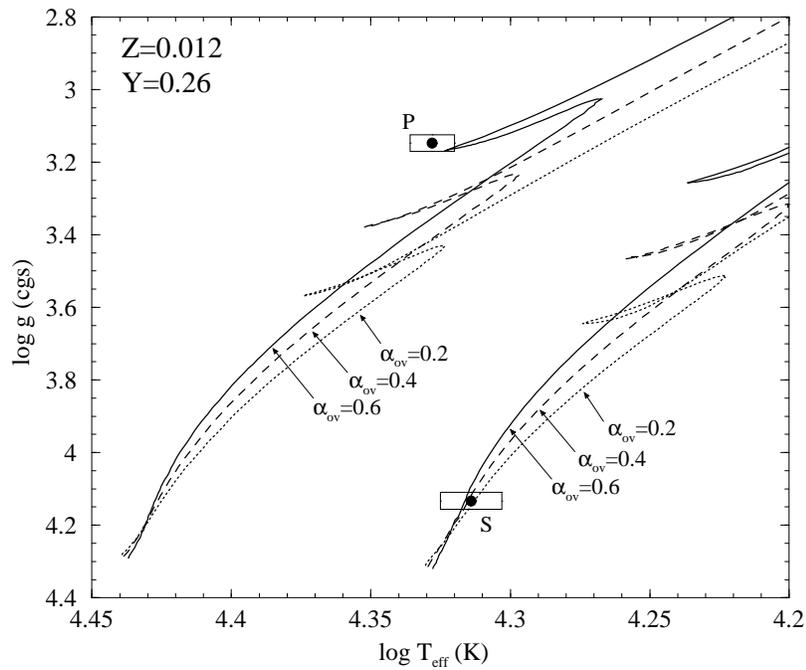}
\figcaption[fig5.eps]{$\log g - \log T_{\rm eff}$ plot of V380~Cyg. 
Evolutionary tracks for the primary (P) and secondary (S) components computed 
with $\alpha_{\rm ov}=0.2$, $0.4$, and $0.6$ are shown. \label{fig:hr}}
\end{figure*}

Evolutionary tracks (specifically generated for V380~Cyg) for three different 
overshooting parameters ($\alpha_{\rm ov}=0.2$, $0.4$, and $0.6$) were kindly 
provided by A. Claret, for the comparison of the observed stellar properties 
with the theoretical predictions. The input physics for these models is
identical (except for the enhanced mixing) to that described in Claret (1995),
i.e., OPAL opacities, mass loss beyond 10~M$_{\odot}$, mixing length parameter
$\alpha_{\rm p}=1.52$, modern equations of state, extensive nuclear network, 
etc. The initial masses of the evolutionary tracks were selected so that they 
would agree with the observed masses of the components at their current 
evolution stage. A metal abundance of $Z=0.012$ was adopted as discussed in \S 
\ref{sec:teff}. The helium abundance ($Y$) that yields the best match to the 
observed properties of the essentially unevolved secondary component (not 
affected by the value of the overshooting parameter, see below) was found to 
be $Y=0.26\pm0.02$.  This value is in good agreement with that expected from 
the metal abundance of V380~Cyg and an average helium-to-metal enrichment law 
(see Ribas et al.  2000a). A $\log T_{\rm eff}-\log g$ plot showing the system 
components and the tracks computed for the adopted chemical composition is 
presented in Figure \ref{fig:hr}.

\placefigure{fig:hr}

The effect of adopting different values of the overshooting parameter
in the early post-ZAMS evolutionary stage of the secondary component is very 
small. Therefore, its observed properties can be well reproduced by models 
with wide ranges of convective overshooting. The evolved primary component, 
however, with $\log g = 3.15$, presents a quite different situation. This star
provides an excellent test of convective overshooting. Indeed, the evolutionary 
track with $\alpha_{\rm ov}=0.2$ predicts an effective temperature for the 
surface gravity of the primary component that is $\sim3000$~K ($\sim7\sigma$) 
lower than the observed value. Even with $\alpha_{\rm ov}=0.4$, the star is 
predicted to have an effective temperature 2000~K lower than observed. A value 
of $\alpha_{\rm ov}=0.6$ is required to achieve consistency between model 
predictions and observations, as shown in Figure \ref{fig:hr}. In this case, 
the primary component is predicted to be close to the phase of hydrogen shell 
ignition. 

Although not explicitly shown in Figure \ref{fig:hr} for the sake of clarity, 
evolutionary tracks for the observed mass plus and minus 1 $\sigma$ were also 
available. The $M + 1 \sigma$ model for the primary component yields effective
temperatures at a given $\log g$ only about 0.009~dex ($\approx 500$~K) larger 
than the model for the observed mass. Therefore, even when considering the 
error bars, the location of the primary component is not compatible with 
evolutionary tracks including mild overshooting. A range of overshooting 
parameters within $\alpha_{\rm ov}\approx0.6\pm0.1$ is possible when all the
observational errors (mass, $\log g$, chemical composition, temperature  
ratio, etc) are taken into account in the analysis.

Even though the effective temperature is a well-determined quantity
in our analysis, it might still be affected by systematics in the model
atmosphere fluxes or in the IUE observations. However, our result favoring a 
large amount of convective core overshooting is immune to any of these 
effects, because the temperature ratio of the components is accurately 
determined from the light curve analysis. The good agreement obtained for the 
secondary component excludes the possibility of resorting to a different value 
of $Y$ or an error in the zero-point of the temperature scale for explaining 
the temperature difference. The conclusion of the analysis is based on the 
tightly-constrained {\em relative} location of the components in the $\log g-
\log T_{\rm eff}$ diagram (i.e., effective temperature ratio) rather than on 
the {\em absolute} value adopted for the effective temperatures. 

It is important to note that in a detached system such as V380~Cyg the ages 
predicted by the models for the two components should agree. This restriction 
is not very critical for V380~Cyg, however, because the secondary component 
is close to the ZAMS and, since the
observational errors in both the mass and the radius is of the order of 2-3 
percent, a wide age range is compatible with the stellar parameters. This
is shown in Table \ref{tab:mod}, where the model predictions (and the
errors) are given for the three overshooting values considered. The 
age, effective temperature, and surface gravity for the primary component
are presented in the table. Also included is the age of the secondary 
component. These
parameters were interpolated in the computed tracks at the observed values of
$\log g$ for both components. The evolutionary ages for the primary and 
secondary components are identical within the errors and no significant 
age-dependent variations are apparent for the various values of the 
overshooting parameter. Therefore, the coeval criterion for the stars does not 
favor any particular overshooting parameter but it confirms the good agreement 
between observed and model predicted physical properties. The age of the 
system was found to be $\tau=25.5\pm1.5$~Myr when the best-fitting 
$\alpha_{\rm ov}\approx 0.6$ was adopted.

A detailed analysis was also carried out with evolutionary tracks of solar 
metal content ($Z=0.02$, $Y=0.28$). From the locations of the stars in the 
$\log g-\log T_{\rm eff}$ we found that $\alpha_{\rm ov}\approx0.6$ still 
clearly yields the best fit, although the components are not compatible with 
coeval evolution (a small age differential of $\sim3$~Myr between the 
components becomes evident).

It has been suggested that the effects of rotation on the structure and
evolution of massive stars could mimic those of overshooting. We used
evolutionary tracks for V380~Cyg provided by A. Claret (the input physics 
and approximations are described in detail in Claret 1999) to explore this 
possibility. Models that include rotation do not show significant 
differences in the $\log T_{\rm eff} -\log g$ diagram when compared 
to the non-rotating models, especially in the evolved stages (a similar 
conclusion was reached by Deupree 1998). Thus, it appears that this effect 
cannot mitigate the relatively high value of the overshooting parameter 
found here. 

In practice, the inclusion of rotation in the models only has significant 
effect when the rotational velocity of the star is fast and 
approaches the critical value (break up). However, V380~Cyg is a slow rotator 
($P\approx12$~days). The initially faster-rotating eclipsing binary 
components are braked by tidal effects until they become synchronized 
with the orbital period. Because the latter is usually quite long for detached 
massive or intermediate-mass systems, the components spin more slowly than 
single stars of the same mass. Therefore, evolutionary models that do not 
include rotation can more confidently be used in eclipsing binaries with
massive components than with more rapidly rotating single stars (in clusters, 
for example).

\placetable{tab:mod}

The determination of the apsidal motion rate of V380~Cyg permits an independent 
test of stellar model predictions. The apsidal motion rate of an eccentric 
binary system can be computed from theory as the sum of the contribution of 
general relativity (GR) and classical (CL) effects (see e.g., Guinan \& Maloney
1985). In a system such as V380~Cyg, with a distorted component, the CL 
contribution to the apsidal motion is expected to dominate the GR term. The 
classical apsidal motion rate is a function of the orbital ($P$, $e$) and 
physical ($M_{\rm P,S}$, $r_{\rm P,S}$, ${v_{\rm rot}}_{\rm P,S}$) properties 
of the system, all determined from the analysis of the light and radial 
velocity curves. But in addition, the classical apsidal motion depends on 
the internal mass distribution of the stars. A star's internal mass
distribution is parameterized in polytropic models by the internal
structure constant $k_2$. In this theory, $k_2$ is proportional to the ratio
of the mean and the central densities of the star. The values of $k_2$
can be computed from stellar evolution codes (see Claret \& Gim\'enez 1992).

An observed apsidal motion rate of $\dot{\omega} {\rm (obs)}= 
24.0\pm1.8$~$^{\circ}$/100~yr was obtained in \S \ref{sec:aps}. For comparison 
with the model predictions, $k_2$ can be computed from $\dot{\omega} {\rm 
(obs)}$ (after correcting for the general relativistic effect, 
${\dot{\omega}}_{\rm GR}= 2$~$^{\circ}$/100~yr) by using the expressions in 
Claret \& Gim\'enez (1993) and by assuming pseudo-synchronization. 
Additionally, several studies (e.g., Stothers 1974; Claret \& Gim\'enez 1993) 
indicate that a small correction may be needed to include the effects of 
stellar rotation in the calculation of $k_2$.

Taking these effects into consideration, we computed a systemic mean 
observational value of $\log {\overline k_2} {\rm (obs)} = -2.89\pm0.04$. 
The contribution of the secondary component to the mean internal structure 
constant is negligible (less than 1 percent) because of its small fractional 
radius. Thus, we assume $\log {k_2}_{\rm P} {\rm (obs)} = -2.89\pm0.05$. The 
specially constructed evolutionary models for V380~Cyg, kindly provided by 
A. Claret, yield the values listed in Table \ref{tab:mod}. As can be seen, 
a value of $\alpha_{\rm ov}$ as large as $0.6\pm0.1$ is clearly needed to fit 
the observed apsidal motion rate of the system.

It is remarkable that the overshooting parameter of $\alpha_{\rm ov} \approx
0.6\pm0.1$ is supported by two independent approaches, namely the investigation
of the apsidal motion and the location of the components in the theoretical
H-R diagram. The value indicated is significantly larger than that currently 
adopted by most theoretical models. Some brief comments on the physical 
implications of this result are left for \S \ref{sec:conc}. In addition, a 
more extensive study on stellar core convection using all the available data 
from eclipsing binaries is the subject of a forthcoming paper (Ribas, Jordi, 
\& Gim\'enez 2000b).

\section{Tidal Evolution} \label{sec:tidal}

V380~Cyg has an eccentric orbit and the component stars appear to have 
rotational velocities synchronized with the orbital velocity at periastron,
the so called pseudo-synchronization (see Table \ref{tab:com}). 
Pseudo-synchronization is commonly found in eccentric binary systems because 
in most cases the circularization timescale is much larger than the 
synchronization timescale (Claret, Gim\'enez, \& Cunha 1995). To compare the 
observations with the theoretical predictions we employed the tidal evolution 
formalism of Tassoul (1987, 1988). The expressions used, 
which were taken from Claret et al. (1995), relate the time variation of the 
eccentricity ($\dot{e}$) and the angular rotation of the components 
($\dot{\Omega}$) with the orbital and stellar physical properties (mass ratio, 
orbital period, mass, luminosity, radius and gyration radius). It is important 
to integrate the differential equation that governs the variation of the 
eccentricity and the angular velocity with time rather than use a simple 
timescale calculation. This is especially advisable for evolved systems like 
V380~Cyg because the components may have undergone large radius and 
luminosity changes that have an important effect over the circularization and 
synchronization timescales.

Thus, we integrated the differential equations along evolutionary tracks 
adopting the observed masses of the V380~Cyg components. In these calculations 
we assumed a constant orbital period. As boundary conditions, we assumed the 
eccentricity to be the observed value ($e=0.234$) at the present age 
of the system ($\tau=25.5$~Myr), and the angular rotation of the stars to be 
normalized to unity at the ZAMS. The calculations were made with the 
evolutionary models at $Z=0.012$, $Y=0.26$, and $\alpha_{\rm ov}=0.6$. The 
predictions of theory agree well with a system that has reached synchronism for 
both components but that has not yet circularized its orbit. These calculations
indicate that the primary component reduced the difference between its 
angular velocity and the pseudo-synchronization angular velocity to $\approx
0.1$ percent of the ZAMS value at the early age of $\approx10$~Myr, and the 
same occurred to the secondary component at an age of $\approx20$~Myr. 

From the boundary condition it is inferred that the initial (ZAMS) orbital 
eccentricity was $e\approx0.3$. A comparison of this initial value with the
current observed one indicates a net decrease of $\Delta e \approx 0.07$ in
about 25~Myr. The theory also predicts that circularization should occur 
on a relatively short timescale of $\approx1.5\times10^4$~yr when the radius
of the primary star rapidly increases shortly before the initiation of 
core helium burning. The instantaneous (i.e., with the current parameters) 
circularization timescale is $\approx10^8$~yr, much larger than the value 
computed from the integration of the differential equation because the 
primary component is located in a very fast evolving stage. Although tidal 
evolution theory predicts orbital circularization for V380~Cyg in 
$1.5\times10^4$~yr, the Roche lobe constraints are not included in the 
modeling. If these are included, the primary star should fill its Roche lobe 
in about $10^4$~yr. This results in Roche lobe overflow and the initiation 
of mass transfer. Once V380~Cyg reaches the semi-detached stage, the binary 
should circularize very rapidly due to strong tidal interactions and friction.

\section{Conclusions} \label{sec:conc}

This study complements the spectroscopic investigation of V380~Cyg published 
in Paper I. New differential photometry has been collected in two different 
epochs, resulting in about 300 observations in each of the $UBV$ bandpasses. 
The photometry has been modeled using an improved version of the 
Wilson-Devinney program that includes recent model atmospheres. The best 
fit to the light curves was achieved for the stellar and orbital properties 
listed in Table \ref{tab:prop}. The effective temperatures of the stars  
were determined using UV spectrophotometry and optical photometry. Also, four 
new eclipse timings were added to the published ones and a complete 
re-evaluation of the apsidal motion rate of V380~Cyg was carried out.

The determined physical properties (Table \ref{tab:com}) of the V380~Cyg 
components were compared with evolutionary models (with the same input 
physics as in Claret 1995) computed for the observed masses, and with 
different values of convective overshoot ($\alpha_{\rm ov}=0.2$, $0.4$, 
and $0.6$). The fractional abundances of metals and helium were estimated to be 
$Z=0.012\pm0.003$ (from UV/optical spectrophotometry) and $Y=0.26\pm0.02$ 
(from evolutionary model fit to the secondary component) respectively. 
The best agreement was found for an overshooting parameter of $\alpha_{\rm ov}
\approx0.6$, for which the primary component is predicted to be located near 
the blue point of the MS hook. The excellent agreement obtained for the 
secondary component excludes the 
possibility of resorting to a different $Y$ value or an error in the 
temperature scale zero point for explaining the location of the primary 
component, because the temperature ratio of the components is well-constrained 
from the light curve analysis. The apsidal motion study of V380~Cyg also 
indicates that a value of $\alpha_{\rm ov}\approx0.6$ is necessary 
to fit the observed apsidal motion rate of the system. The study of the 
locations of the components in the $\log g - \log T_{\rm eff}$ diagram and 
the internal structure (from apsidal motion) lead to mutually consistent 
results. However, we caution that the relatively large value of overshooting 
should be restricted to the mass, evolutionary stage, and chemical composition 
of the primary component of V380~Cyg.

%The result of our study is more profound than simply a larger 
%overshooting parameter than currently adopted. We have shown that, in the
%absence of any degree of freedom, the physical properties of V380~Cyg 
%cannot be described by evolutionary models with the most commonly used 
%physical ingredients (including mild overshooting). In an attempt to explain 
%such discrepancy, we have included a free parameter related to the convection 
%model, namely core overshooting. When adopting this parameterization, we have 
%found that the agreement between observations and theory 
%is extraordinarily improved when increasing the amount of overshooting. In a 
%more general and physical sense: massive stars seem to have larger convective 
%cores and, thus, to be more centrally condensed than predicted by standard 
%theory. Not only a larger overshooting parameter, however, explains such 
%observational fact. 

Our convective overshooting result has a more profound implication than
merely providing a better fit to the V380~Cyg data. We have shown that the 
observed physical properties of V380~Cyg cannot be described by evolutionary 
models with the most commonly used physical ingredients (including mild 
overshooting). The physical difference between these standard models and 
our better-fitting models (with enhanced convective overshooting)
is that the latter have larger convective cores. Thus, our more general 
conclusion is that massive stars have larger convective cores and, therefore, 
are more centrally condensed than predicted by standard theory. Note that 
convective overshooting is not the only physical process that might produce 
such a result. Further increases in the opacity (as it was demonstrated 
when substituting the LAOL by the newer OPAL opacity tables), diffusive
mixing, turbulence or other mechanisms may certainly lead to stellar
models with larger and denser convective cores, relaxing the need for our 
large overshooting parameter. In the meantime, with the current physical
ingredients, our results indicate that models should consider a moderate 
amount of convective overshooting for matching the observed properties of 
real massive stars.

An investigation of the tidal evolution of V380~Cyg was also carried out. The 
numerical integrations of the relevant differential equations indicate a binary 
system with components in an eccentric orbit that have reached 
pseudo-synchronism.  From the theory of Tassoul (1987, 1988), it appears that 
the orbit circularization should occur on a timescale of $10^4$ years. This 
is on the same timescale as it takes the primary component to fill its Roche 
lobe and start mass transfer. 

This paper clearly demonstrates the importance of specific eclipsing binaries
as ``astrophysical laboratories''. In the present case of V380~Cyg,
fundamental problems of stellar structure and evolution are addressed. 
Early-type massive systems with at least one evolved component constitute a 
very important source of observational data for testing stellar structure and 
core convection. Such systems are scarce (due to strong observational selection 
effects) and gathering the required photometric and spectroscopic data is not 
easy (because of the typically long orbital periods -- 10 days or more). The 
relative faintness of the secondary component also presents difficulties in 
the spectroscopic aspects of the problem. In spite of the observational 
challenges, these systems (although rare) have proved to be crucial tools 
and their study should be a priority in stellar astrophysics. 

Some examples of early-type eclipsing binaries suitable for testing core 
convection are: V1756~Cyg, V453~Cyg and V346~Cen (see Ribas et al. 1999). 
V1765~Cyg has a very massive and evolved component, already in the supergiant
stage. V453~Cyg and V346~Cen have primary components of similar mass to that 
of V380~Cyg but appear to be at a less evolved stage ($\log g \approx 3.8$). 
Other suitable binaries for testing models are those belonging to the 
$\zeta$-Aur 
class of eclipsing binaries (Schr\"oder, Pols, \& Eggleton 1997). These are 
systems with at least one evolved member in the core helium burning or 
in the supergiant stage. The long periods (hundreds or even thousands of days) 
make the observations challenging. Hopefully, there will be many more 
potentially interesting systems to study. The long term monitoring programs 
associated with (future) space missions and the ground-based systematic surveys 
(e.g., the Sloan Digital Survey and the OGLE experiment towards the Galactic 
Bulge) are expected to yield valuable results, both from the detection point of 
view and from the densely-covered light curves that they will provide. Finally, 
extensive photometry of extragalactic eclipsing binaries (in LMC, SMC, and 
M31) are becoming available as a result of several microlensing 
(Grison et al. 1995; Alcock et al. 1997; Udalski et al. 1998) and variable 
star (Kaluzny et al. 1998, 1999; Stanek et al. 1998, 1999) surveys. Such a 
wealth of data guarantees a nearly inexhaustible source of eclipsing systems 
meeting the demanding requirements for critical model analysis, not only for 
solar-type abundances but also for other chemical compositions.

\acknowledgments
This paper is dedicated to the memory of Dr. Daniel M. Popper who died last 
September before this work could be completed. We are grateful for his 
inspiration and for his major contributions to eclipsing binaries. Dr. Antonio
Claret is gratefully thanked for computing specific evolutionary tracks for 
this study. We also wish to thank the contributions of Bryan Deeney and
Dr. David H. Bradstreet for help during the early stages of the project. The
referee, Dr. Andy Odell, is thanked for a number of important comments and 
suggestions that 
led to the improvement of the paper. This work was supported by NASA grant 
NAG5-2160 and NSF/RUI grants AST 93-15365 and AST 95-28506. I. R. acknowledges 
support from the predoctoral grant (ref.  FI-PG/95-1111) and the postdoctoral 
Fulbright fellowship by the Catalan Regional Government (CIRIT). E. L. F. 
acknowledges support from NASA ADP grant NAG5-7117 to Villanova University.

%
%  Begin Bibliography Page
%

%  Begin Figure Caption Page

\clearpage

%  Begin Table Pages

\clearpage

\begin{deluxetable}{rrrrrrrrrrrr}
\tabletypesize{\tiny}
\tablewidth{0pt}
\tablecaption{Differential photometry for V380~Cyg in the $UBV$ bandpasses.
\label{tab:phot}}
\tablehead{\colhead{HJD$-$}& \colhead{V-C}&
\colhead{HJD$-$}& \colhead{V$-$C}&
\colhead{HJD$-$}& \colhead{V$-$C}&
\colhead{HJD$-$}& \colhead{V$-$C}&
\colhead{HJD$-$}& \colhead{V$-$C}&
\colhead{HJD$-$}& \colhead{V$-$C}\\
\colhead{2400000}& \colhead{}&
\colhead{2400000}& \colhead{}&
\colhead{2400000}& \colhead{}&
\colhead{2400000}& \colhead{}&
\colhead{2400000}& \colhead{}&
\colhead{2400000}& \colhead{}}
\startdata
\multicolumn{12}{c}{$U$}\\
\tableline
 47415.8013 &$-$0.004 & 47417.7500 &   0.003 & 47418.7531 &   0.010 & 47419.7573 &   0.107 & 47420.7519 &   0.007 & 47421.7429 &$-$0.023 \\
 47422.7256 &$-$0.003 & 47423.7161 &   0.015 & 47424.7076 &   0.081 & 47427.7040 &$-$0.009 & 47428.7033 &$-$0.007 & 47429.7106 &   0.004 \\
 47430.7018 &   0.005 & 47431.7010 &   0.042 & 47432.7083 &   0.023 & 47433.7389 &$-$0.030 & 47436.7233 &   0.048 & 47437.7146 &   0.012 \\
 47441.7282 &$-$0.013 & 47443.7104 &   0.002 & 47444.7108 &   0.081 & 47459.6380 &$-$0.017 & 47460.6368 &   0.004 & 47461.6359 &   0.075 \\
 47463.6422 &   0.005 & 47465.6435 &$-$0.003 & 47466.6301 &$-$0.015 & 47468.6640 &   0.019 & 47469.6220 &   0.097 & 47470.6211 &$-$0.021 \\
 47472.6245 &$-$0.007 & 47475.6085 &   0.005 & 47476.6031 &   0.004 & 47479.5852 &$-$0.021 & 47481.5778 &   0.081 & 47482.5705 &   0.006 \\
 47483.5706 &$-$0.035 & 47485.5705 &$-$0.014 & 47486.5703 &   0.057 & 47494.5554 &   0.058 & 47495.5612 &$-$0.022 & 47497.5610 &$-$0.007 \\
 47498.5546 &   0.004 & 47500.5611 &$-$0.007 & 47502.5549 &   0.004 & 47507.5554 &   0.003 & 47618.0173 &$-$0.004 & 47619.0159 &   0.016 \\
 47621.0149 &$-$0.029 & 47622.0135 &$-$0.012 & 47623.0106 &   0.032 & 47626.9985 &$-$0.027 & 47629.9922 &$-$0.010 & 47631.9828 &$-$0.005 \\
 47632.9770 &$-$0.037 & 47635.9776 &   0.058 & 47636.9714 &$-$0.009 & 47641.9575 &$-$0.020 & 47646.9470 &$-$0.002 & 47647.9447 &   0.039 \\
 47648.9401 &$-$0.004 & 47650.9327 &   0.001 & 47651.9321 &$-$0.019 & 47664.8964 &$-$0.021 & 47665.8909 &$-$0.023 & 47668.8804 &   0.010 \\
 47669.9391 &$-$0.021 & 47670.8765 &$-$0.020 & 47671.9653 &$-$0.008 & 47672.8840 &   0.048 & 47673.8940 &$-$0.001 & 47680.8474 &   0.077 \\
 47683.8784 &$-$0.017 & 47686.8887 &   0.006 & 47688.8657 &$-$0.011 & 47689.8647 &$-$0.013 & 47690.8620 &$-$0.021 & 47691.8520 &$-$0.008 \\
 47693.8433 &   0.004 & 47694.8597 &$-$0.025 & 47695.8595 &$-$0.022 & 47699.8585 &   0.008 & 47701.8618 &$-$0.018 & 49864.9458 &$-$0.023 \\
 49864.9472 &$-$0.012 & 49864.9482 &$-$0.022 & 49864.9496 &$-$0.017 & 49864.9511 &$-$0.012 & 49866.9329 &$-$0.014 & 49866.9343 &   0.007 \\
 49866.9353 &   0.008 & 49866.9367 &   0.009 & 49866.9382 &   0.016 & 49867.9052 &   0.052 & 49867.9066 &   0.085 & 49867.9076 &   0.073 \\
 49867.9090 &   0.070 & 49867.9105 &   0.066 & 49868.9408 &$-$0.023 & 49868.9422 &$-$0.013 & 49868.9431 &$-$0.022 & 49868.9446 &$-$0.008 \\
 49868.9460 &$-$0.015 & 49869.9487 &$-$0.030 & 49869.9511 &$-$0.034 & 49869.9526 &$-$0.037 & 49870.9376 &$-$0.027 & 49870.9390 &   0.011 \\
 49870.9400 &$-$0.018 & 49870.9414 &   0.004 & 49870.9429 &$-$0.025 & 49871.9413 &   0.005 & 49871.9427 &   0.001 & 49871.9442 &$-$0.001 \\
 49872.9382 &   0.025 & 49872.9396 &   0.016 & 49872.9406 &   0.013 & 49872.9435 &   0.032 & 49873.9213 &   0.017 & 49873.9223 &   0.014 \\
 49873.9237 &   0.015 & 49873.9252 &   0.004 & 49878.9438 &$-$0.011 & 49878.9448 &   0.004 & 49878.9477 &$-$0.018 & 49879.9263 &   0.085 \\
 49879.9277 &   0.104 & 49879.9287 &   0.094 & 49879.9301 &   0.093 & 49879.9316 &   0.073 & 49880.9014 &$-$0.008 & 49880.9028 &$-$0.006 \\
 49880.9037 &   0.007 & 49880.9052 &   0.026 & 49880.9067 &   0.001 & 49883.9255 &$-$0.011 & 49883.9269 &$-$0.003 & 49883.9279 &   0.003 \\
 49883.9293 &   0.003 & 49883.9308 &$-$0.011 & 49884.8753 &   0.072 & 49884.8767 &   0.069 & 49884.8777 &   0.051 & 49884.8791 &   0.078 \\
 49884.8806 &   0.078 & 49885.9681 &$-$0.009 & 49885.9696 &   0.005 & 49885.9705 &$-$0.001 & 49885.9720 &   0.011 & 49885.9735 &   0.006 \\
 49886.9110 &$-$0.008 & 49886.9124 &$-$0.004 & 49886.9134 &   0.008 & 49886.9148 &   0.001 & 49886.9163 &$-$0.008 & 49887.8951 &$-$0.018 \\
 49887.8966 &$-$0.006 & 49887.8976 &$-$0.006 & 49887.8990 &$-$0.001 & 49888.9279 &$-$0.019 & 49888.9293 &$-$0.014 & 49888.9303 &$-$0.018 \\
 49888.9332 &$-$0.016 & 49889.9547 &$-$0.009 & 49889.9556 &$-$0.012 & 49889.9571 &$-$0.006 & 49889.9586 &$-$0.007 & 49890.8945 &   0.005 \\
 49890.8955 &$-$0.001 & 49890.8970 &$-$0.007 & 49890.8984 &$-$0.007 & 49891.8914 &   0.017 & 49891.8924 &   0.018 & 49891.8938 &   0.034 \\
 49891.8953 &   0.015 & 49892.8909 &   0.033 & 49892.8918 &   0.014 & 49892.8933 &   0.031 & 49892.8947 &   0.023 & 49893.8813 &$-$0.022 \\
 49893.8827 &$-$0.005 & 49893.8837 &$-$0.032 & 49893.8851 &$-$0.015 & 49893.8866 &$-$0.022 & 49895.8368 &$-$0.001 & 49895.8377 &$-$0.007 \\
 49895.8392 &   0.007 & 49895.8406 &$-$0.008 & 49896.8944 &   0.042 & 49896.8958 &   0.044 & 49896.8969 &   0.042 & 49896.8983 &   0.059 \\
 49896.8998 &   0.042 & 49898.8474 &$-$0.009 & 49898.8488 &   0.006 & 49898.8498 &   0.009 & 49898.8512 &   0.007 & 49898.8527 &$-$0.002 \\
 49899.8733 &$-$0.020 & 49899.8747 &   0.011 & 49899.8757 &   0.001 & 49899.8771 &$-$0.005 & 49899.8786 &$-$0.006 & 49900.8427 &$-$0.010 \\
 49900.8441 &   0.002 & 49901.8802 &$-$0.025 & 49901.8816 &$-$0.023 & 49901.8826 &$-$0.021 & 49901.8840 &$-$0.005 & 49901.8855 &   0.004 \\
 50237.9720 &$-$0.014 & 50237.9734 &$-$0.002 & 50237.9749 &$-$0.005 & 50251.9572 &$-$0.016 & 50251.9587 &   0.006 & 50251.9602 &   0.002 \\
 50252.9618 &   0.079 & 50252.9632 &   0.105 & 50252.9647 &   0.098 & 50253.9527 &$-$0.023 & 50253.9542 &$-$0.018 & 50253.9557 &$-$0.018 \\
 50254.9480 &$-$0.023 & 50254.9495 &$-$0.008 & 50254.9510 &$-$0.039 & 50259.9444 &$-$0.007 & 50259.9459 &   0.006 & 50259.9473 &$-$0.001 \\
 50396.5737 &   0.000 & 50396.5752 &   0.014 & 50396.5767 &   0.001 & 50714.6782 &$-$0.036 & 50714.6797 &$-$0.036 & 50718.7014 &   0.014 \\
 50718.7028 &$-$0.008 & 50725.6385 &   0.000 & 50725.6399 &   0.028 & 50725.6415 &   0.021 & 50726.6660 &$-$0.033 & 50726.6674 &$-$0.011 \\
 50726.6690 &$-$0.023 & 50727.6718 &$-$0.031 & 50727.6733 &$-$0.025 & 50727.6748 &$-$0.007 & 50728.6355 &   0.014 & 50728.6370 &$-$0.011 \\
 50730.6641 &   0.015 & 50730.6656 &   0.014 & 50730.6671 &   0.018 & 50752.5805 &   0.004 & 50752.5820 &$-$0.026 & 50752.5835 &$-$0.014 \\
 50753.5971 &   0.018 & 50753.5986 &   0.022 & 50753.6001 &$-$0.002 & 50754.5965 &   0.079 & 50754.5980 &   0.081 & 50754.5995 &   0.078 \\
 50755.5957 &   0.007 & 50755.5971 &   0.007 & 50755.5987 &   0.019 & 50756.5951 &   0.002 & 50756.5966 &   0.002 & 50756.5981 &   0.016 \\
\tableline
\multicolumn{12}{c}{$B$}\\
\tableline
 47418.7531 &   0.260 & 47419.7573 &   0.349 & 47420.7519 &   0.253 & 47421.7429 &   0.228 & 47422.7256 &   0.250 & 47423.7161 &   0.261 \\
 47424.7076 &   0.332 & 47427.7040 &   0.240 & 47428.7033 &   0.239 & 47429.7106 &   0.236 & 47430.7018 &   0.259 & 47431.7010 &   0.294 \\
 47432.7083 &   0.275 & 47433.7389 &   0.225 & 47436.7233 &   0.303 & 47437.7146 &   0.270 & 47441.7282 &   0.247 & 47443.7104 &   0.255 \\
 47444.7108 &   0.331 & 47459.6380 &   0.240 & 47460.6368 &   0.262 & 47461.6359 &   0.327 & 47463.6422 &   0.258 & 47465.6435 &   0.252 \\
 47466.6301 &   0.241 & 47469.6220 &   0.349 & 47470.6211 &   0.233 & 47472.6245 &   0.249 & 47475.6085 &   0.259 & 47476.6031 &   0.259 \\
 47479.5852 &   0.237 & 47481.5778 &   0.345 & 47482.5705 &   0.266 & 47483.5706 &   0.232 & 47485.5705 &   0.249 & 47486.5703 &   0.323 \\
 47493.5622 &   0.271 & 47494.5554 &   0.326 & 47495.5612 &   0.232 & 47497.5610 &   0.255 & 47498.5546 &   0.259 & 47500.5611 &   0.251 \\
 47502.5549 &   0.244 & 47507.5554 &   0.278 & 47508.5560 &   0.228 & 47618.0173 &   0.281 & 47619.0159 &   0.285 & 47620.0132 &   0.224 \\
 47621.0149 &   0.244 & 47623.0106 &   0.292 & 47624.0024 &   0.272 & 47626.9985 &   0.239 & 47629.9922 &   0.250 & 47631.9828 &   0.251 \\
 47632.9770 &   0.226 & 47633.9797 &   0.234 & 47635.9776 &   0.328 & 47636.9714 &   0.252 & 47646.9470 &   0.260 & 47647.9447 &   0.297 \\
 47648.9401 &   0.259 & 47649.9361 &   0.249 & 47650.9327 &   0.255 & 47651.9321 &   0.248 & 47664.8964 &   0.243 & 47665.8909 &   0.241 \\
 47668.8804 &   0.271 & 47669.9391 &   0.238 & 47670.8765 &   0.243 & 47671.9653 &   0.261 & 47672.8840 &   0.315 & 47673.8940 &   0.259 \\
 47680.8474 &   0.336 & 47683.8784 &   0.249 & 47686.8887 &   0.251 & 47688.8657 &   0.251 & 47689.8647 &   0.250 & 47690.8620 &   0.245 \\
 47691.8520 &   0.248 & 47693.8433 &   0.257 & 47695.8595 &   0.247 & 47699.8585 &   0.261 & 47701.8618 &   0.247 & 47779.7548 &   0.331 \\
 47780.7559 &   0.268 & 47786.7443 &   0.263 & 47788.7323 &   0.241 & 47793.7278 &   0.244 & 47794.7533 &   0.239 & 47797.7467 &   0.328 \\
 47799.7394 &   0.248 & 47800.7425 &   0.243 & 47807.6961 &   0.247 & 47812.6939 &   0.246 & 47813.6851 &   0.241 & 47816.6890 &   0.256 \\
 47817.6812 &   0.294 & 47818.6881 &   0.233 & 47826.6739 &   0.239 & 47833.6399 &   0.257 & 47834.6378 &   0.331 & 47835.6322 &   0.253 \\
 47836.6230 &   0.255 & 47838.6264 &   0.252 & 47840.6143 &   0.245 & 49864.9459 &   0.240 & 49864.9473 &   0.254 & 49864.9483 &   0.247 \\
 49864.9498 &   0.248 & 49864.9512 &   0.243 & 49866.9330 &   0.265 & 49866.9344 &   0.277 & 49866.9354 &   0.268 & 49866.9368 &   0.271 \\
 49866.9383 &   0.269 & 49867.9053 &   0.321 & 49867.9067 &   0.324 & 49867.9077 &   0.326 & 49867.9091 &   0.325 & 49867.9106 &   0.323 \\
 49868.9409 &   0.241 & 49868.9423 &   0.252 & 49868.9433 &   0.245 & 49868.9447 &   0.246 & 49868.9462 &   0.244 & 49869.9474 &   0.235 \\
 49869.9498 &   0.238 & 49869.9528 &   0.240 & 49870.9377 &   0.251 & 49870.9391 &   0.255 & 49870.9401 &   0.253 & 49870.9416 &   0.255 \\
 49870.9430 &   0.253 & 49871.9390 &   0.258 & 49871.9405 &   0.268 & 49871.9414 &   0.267 & 49871.9428 &   0.271 & 49871.9443 &   0.265 \\
 49872.9383 &   0.294 & 49872.9397 &   0.299 & 49872.9408 &   0.294 & 49872.9422 &   0.301 & 49872.9436 &   0.297 & 49873.9200 &   0.254 \\
 49873.9215 &   0.265 & 49873.9224 &   0.259 & 49873.9239 &   0.259 & 49873.9253 &   0.259 & 49876.8984 &   0.239 & 49876.8999 &   0.243 \\
 49876.9023 &   0.241 & 49878.9425 &   0.261 & 49878.9440 &   0.260 & 49878.9449 &   0.258 & 49878.9464 &   0.265 & 49878.9478 &   0.261 \\
 49879.9264 &   0.346 & 49879.9279 &   0.353 & 49879.9288 &   0.347 & 49879.9303 &   0.347 & 49879.9318 &   0.344 & 49880.9015 &   0.261 \\
 49880.9029 &   0.273 & 49880.9039 &   0.265 & 49880.9053 &   0.265 & 49880.9068 &   0.262 & 49883.9257 &   0.251 & 49883.9271 &   0.263 \\
 49883.9281 &   0.260 & 49883.9295 &   0.266 & 49883.9310 &   0.260 & 49884.8754 &   0.329 & 49884.8769 &   0.339 & 49884.8778 &   0.335 \\
 49884.8792 &   0.337 & 49884.8807 &   0.338 & 49885.9683 &   0.254 & 49885.9697 &   0.263 & 49885.9707 &   0.261 & 49885.9721 &   0.263 \\
 49885.9736 &   0.259 & 49886.9111 &   0.249 & 49886.9126 &   0.253 & 49886.9135 &   0.253 & 49886.9149 &   0.261 & 49886.9164 &   0.250 \\
 49887.8953 &   0.242 & 49887.8967 &   0.252 & 49887.8977 &   0.248 & 49887.8991 &   0.255 & 49887.9006 &   0.248 & 49888.9280 &   0.242 \\
 49888.9295 &   0.249 & 49888.9304 &   0.242 & 49888.9319 &   0.251 & 49888.9334 &   0.246 & 49889.9534 &   0.238 & 49889.9548 &   0.243 \\
 49889.9558 &   0.243 & 49889.9572 &   0.246 & 49889.9587 &   0.244 & 49890.8932 &   0.245 & 49890.8947 &   0.253 & 49890.8956 &   0.257 \\
 49890.8971 &   0.254 & 49890.8986 &   0.251 & 49891.8901 &   0.263 & 49891.8915 &   0.273 & 49891.8925 &   0.269 & 49891.8939 &   0.277 \\
 49891.8954 &   0.272 & 49892.8896 &   0.286 & 49892.8910 &   0.295 & 49892.8920 &   0.290 & 49892.8934 &   0.294 & 49892.8949 &   0.291 \\
 49893.8815 &   0.231 & 49893.8829 &   0.242 & 49893.8838 &   0.238 & 49893.8853 &   0.237 & 49893.8868 &   0.236 & 49895.8369 &   0.250 \\
 49895.8379 &   0.243 & 49895.8393 &   0.249 & 49895.8408 &   0.242 & 49896.8945 &   0.298 & 49896.8960 &   0.300 & 49896.8970 &   0.303 \\
 49896.8984 &   0.302 & 49896.8999 &   0.299 & 49898.8476 &   0.248 & 49898.8490 &   0.255 & 49898.8499 &   0.255 & 49898.8514 &   0.254 \\
 49898.8529 &   0.252 & 49899.8734 &   0.246 & 49899.8748 &   0.263 & 49899.8758 &   0.255 & 49899.8772 &   0.260 & 49899.8787 &   0.255 \\
 49900.8428 &   0.243 & 49900.8443 &   0.255 & 49900.8467 &   0.260 & 49901.8803 &   0.239 & 49901.8817 &   0.247 & 49901.8828 &   0.247 \\
 49901.8842 &   0.249 & 49901.8857 &   0.240 & 50237.9721 &   0.242 & 50237.9735 &   0.250 & 50237.9750 &   0.247 & 50251.9588 &   0.258 \\
 50252.9619 &   0.351 & 50252.9649 &   0.348 & 50253.9529 &   0.239 & 50253.9543 &   0.246 & 50253.9558 &   0.257 & 50254.9481 &   0.228 \\
 50254.9496 &   0.235 & 50254.9511 &   0.223 & 50259.9446 &   0.248 & 50259.9460 &   0.258 & 50259.9475 &   0.250 & 50396.5739 &   0.247 \\
 50396.5753 &   0.253 & 50396.5768 &   0.252 & 50713.6780 &   0.252 & 50713.6794 &   0.257 & 50714.6783 &   0.228 & 50714.6798 &   0.236 \\
 50718.7015 &   0.249 & 50718.7029 &   0.259 & 50718.7044 &   0.248 & 50721.6658 &   0.239 & 50721.6672 &   0.254 & 50721.6688 &   0.243 \\
 50725.6386 &   0.266 & 50725.6401 &   0.275 & 50725.6416 &   0.268 & 50726.6661 &   0.223 & 50726.6676 &   0.232 & 50726.6691 &   0.229 \\
 50727.6719 &   0.238 & 50727.6734 &   0.246 & 50727.6749 &   0.240 & 50728.6342 &   0.256 & 50728.6356 &   0.263 & 50728.6372 &   0.259 \\
 50730.6658 &   0.256 & 50752.5807 &   0.237 & 50752.5821 &   0.244 & 50752.5836 &   0.235 & 50753.5973 &   0.257 & 50753.5987 &   0.259 \\
 50753.6002 &   0.250 & 50754.5966 &   0.334 & 50754.5997 &   0.336 & 50755.5958 &   0.258 & 50755.5973 &   0.264 & 50755.5988 &   0.258 \\
 50756.5952 &   0.252 & 50756.5967 &   0.266 & 50756.5982 &   0.257 &    \nodata & \nodata &    \nodata & \nodata &    \nodata & \nodata \\
\tableline
\multicolumn{12}{c}{$V$}\\
\tableline
 47415.8013 &   0.217 & 47417.7500 &   0.204 & 47418.7531 &   0.227 & 47420.7519 &   0.221 & 47421.7429 &   0.196 & 47422.7256 &   0.215 \\
 47423.7161 &   0.229 & 47424.7076 &   0.297 & 47427.7040 &   0.207 & 47428.7033 &   0.208 & 47429.7106 &   0.208 & 47431.7010 &   0.256 \\
 47432.7083 &   0.239 & 47433.7389 &   0.186 & 47436.7233 &   0.268 & 47441.7282 &   0.209 & 47443.7104 &   0.219 & 47459.6380 &   0.204 \\
 47461.6359 &   0.287 & 47466.6301 &   0.207 & 47469.6220 &   0.311 & 47470.6211 &   0.200 & 47472.6245 &   0.217 & 47475.6085 &   0.218 \\
 47476.6031 &   0.221 & 47479.5852 &   0.205 & 47482.5705 &   0.228 & 47483.5706 &   0.194 & 47485.5705 &   0.221 & 47486.5703 &   0.288 \\
 47493.5622 &   0.234 & 47494.5554 &   0.288 & 47495.5612 &   0.201 & 47497.5610 &   0.222 & 47498.5546 &   0.227 & 47500.5611 &   0.216 \\
 47508.5560 &   0.193 & 47618.0173 &   0.247 & 47619.0159 &   0.246 & 47620.0132 &   0.189 & 47622.0135 &   0.225 & 47623.0106 &   0.256 \\
 47624.0024 &   0.231 & 47626.9985 &   0.202 & 47629.9922 &   0.218 & 47631.9828 &   0.215 & 47632.9770 &   0.187 & 47635.9776 &   0.288 \\
 47636.9714 &   0.216 & 47641.9575 &   0.218 & 47646.9470 &   0.226 & 47647.9447 &   0.264 & 47648.9401 &   0.219 & 47649.9361 &   0.223 \\
 47651.9321 &   0.213 & 47664.8964 &   0.212 & 47665.8909 &   0.204 & 47668.8804 &   0.234 & 47669.9391 &   0.194 & 47670.8765 &   0.197 \\
 47671.9653 &   0.225 & 47672.8840 &   0.280 & 47673.8940 &   0.225 & 47677.8624 &   0.203 & 47680.8474 &   0.301 & 47683.8784 &   0.206 \\
 47686.8887 &   0.216 & 47687.8811 &   0.208 & 47688.8657 &   0.199 & 47689.8647 &   0.215 & 47690.8620 &   0.199 & 47691.8520 &   0.216 \\
 47693.8433 &   0.222 & 47694.8597 &   0.191 & 47695.8595 &   0.208 & 47699.8585 &   0.214 & 47701.8618 &   0.207 & 47779.7548 &   0.293 \\
 47780.7559 &   0.232 & 47788.7323 &   0.199 & 47789.7426 &   0.211 & 47793.7278 &   0.208 & 47794.7533 &   0.200 & 47797.7467 &   0.291 \\
 47799.7394 &   0.210 & 47800.7425 &   0.209 & 47812.6939 &   0.206 & 47813.6851 &   0.201 & 47814.6905 &   0.211 & 47816.6890 &   0.216 \\
 47817.6812 &   0.253 & 47818.6881 &   0.198 & 47826.6739 &   0.207 & 47827.6683 &   0.203 & 47833.6399 &   0.221 & 47834.6378 &   0.293 \\
 47835.6322 &   0.214 & 47836.6230 &   0.219 & 47838.6264 &   0.213 & 47840.6143 &   0.209 & 49864.9461 &   0.200 & 49864.9475 &   0.205 \\
 49864.9485 &   0.202 & 49864.9499 &   0.208 & 49864.9514 &   0.203 & 49866.9332 &   0.226 & 49866.9346 &   0.219 & 49866.9356 &   0.224 \\
 49866.9370 &   0.228 & 49866.9385 &   0.221 & 49867.9054 &   0.282 & 49867.9069 &   0.280 & 49867.9079 &   0.279 & 49867.9093 &   0.280 \\
 49867.9108 &   0.279 & 49868.9410 &   0.201 & 49868.9425 &   0.207 & 49868.9434 &   0.207 & 49868.9448 &   0.210 & 49868.9463 &   0.199 \\
 49869.9476 &   0.194 & 49869.9490 &   0.201 & 49869.9500 &   0.196 & 49869.9514 &   0.200 & 49869.9529 &   0.197 & 49870.9379 &   0.213 \\
 49870.9393 &   0.213 & 49870.9403 &   0.214 & 49870.9417 &   0.215 & 49870.9432 &   0.213 & 49871.9392 &   0.224 & 49871.9406 &   0.229 \\
 49871.9416 &   0.222 & 49871.9430 &   0.226 & 49871.9445 &   0.226 & 49872.9385 &   0.246 & 49872.9399 &   0.256 & 49872.9409 &   0.254 \\
 49872.9423 &   0.253 & 49872.9438 &   0.253 & 49873.9201 &   0.217 & 49873.9216 &   0.226 & 49873.9226 &   0.211 & 49873.9240 &   0.221 \\
 49873.9255 &   0.217 & 49876.8986 &   0.201 & 49876.9000 &   0.201 & 49876.9010 &   0.201 & 49876.9024 &   0.205 & 49876.9039 &   0.195 \\
 49878.9427 &   0.216 & 49878.9441 &   0.221 & 49878.9451 &   0.212 & 49878.9465 &   0.212 & 49878.9480 &   0.216 & 49879.9266 &   0.299 \\
 49879.9280 &   0.303 & 49879.9290 &   0.299 & 49879.9304 &   0.307 & 49879.9319 &   0.302 & 49880.9016 &   0.214 & 49880.9031 &   0.222 \\
 49880.9040 &   0.219 & 49880.9055 &   0.221 & 49880.9069 &   0.228 & 49883.9258 &   0.209 & 49883.9272 &   0.219 & 49883.9282 &   0.218 \\
 49883.9296 &   0.224 & 49883.9311 &   0.212 & 49884.8756 &   0.283 & 49884.8770 &   0.290 & 49884.8780 &   0.286 & 49884.8794 &   0.295 \\
 49884.8809 &   0.285 & 49885.9684 &   0.208 & 49885.9699 &   0.217 & 49885.9708 &   0.215 & 49885.9722 &   0.222 & 49885.9737 &   0.217 \\
 49886.9113 &   0.210 & 49886.9127 &   0.212 & 49886.9137 &   0.209 & 49886.9151 &   0.215 & 49886.9166 &   0.208 & 49887.8954 &   0.206 \\
 49887.8968 &   0.210 & 49887.8978 &   0.204 & 49887.8993 &   0.209 & 49887.9007 &   0.203 & 49888.9282 &   0.199 & 49888.9296 &   0.203 \\
 49888.9306 &   0.205 & 49888.9320 &   0.201 & 49888.9335 &   0.199 & 49889.9535 &   0.198 & 49889.9549 &   0.206 & 49889.9559 &   0.202 \\
 49889.9573 &   0.201 & 49889.9588 &   0.200 & 49890.8934 &   0.209 & 49890.8948 &   0.216 & 49890.8958 &   0.207 & 49890.8972 &   0.203 \\
 49890.8987 &   0.201 & 49891.8902 &   0.222 & 49891.8917 &   0.228 & 49891.8927 &   0.221 & 49891.8941 &   0.228 & 49891.8956 &   0.224 \\
 49892.8897 &   0.241 & 49892.8912 &   0.245 & 49892.8921 &   0.246 & 49892.8935 &   0.251 & 49892.8950 &   0.247 & 49893.8816 &   0.193 \\
 49893.8830 &   0.195 & 49893.8840 &   0.193 & 49893.8854 &   0.200 & 49893.8869 &   0.195 & 49895.8356 &   0.206 & 49895.8370 &   0.207 \\
 49895.8380 &   0.203 & 49895.8394 &   0.204 & 49895.8409 &   0.204 & 49896.8946 &   0.253 & 49896.8961 &   0.260 & 49896.8971 &   0.254 \\
 49896.8986 &   0.260 & 49896.9001 &   0.252 & 49898.8477 &   0.204 & 49898.8491 &   0.209 & 49898.8501 &   0.207 & 49898.8515 &   0.208 \\
 49898.8530 &   0.207 & 49899.8736 &   0.202 & 49899.8750 &   0.213 & 49899.8760 &   0.209 & 49899.8774 &   0.211 & 49899.8789 &   0.211 \\
 49900.8430 &   0.208 & 49900.8444 &   0.210 & 49900.8468 &   0.214 & 49901.8804 &   0.196 & 49901.8819 &   0.205 & 49901.8829 &   0.201 \\
 49901.8843 &   0.206 & 49901.8858 &   0.201 & 50237.9722 &   0.205 & 50237.9737 &   0.207 & 50237.9752 &   0.208 & 50251.9590 &   0.213 \\
 50252.9621 &   0.306 & 50252.9635 &   0.315 & 50252.9650 &   0.307 & 50253.9530 &   0.200 & 50253.9544 &   0.213 & 50253.9559 &   0.205 \\
 50254.9483 &   0.184 & 50254.9497 &   0.192 & 50259.9447 &   0.204 & 50259.9462 &   0.215 & 50259.9476 &   0.209 & 50396.5740 &   0.209 \\
 50396.5754 &   0.209 & 50396.5769 &   0.209 & 50713.6796 &   0.212 & 50714.6785 &   0.185 & 50714.6799 &   0.193 & 50718.7016 &   0.204 \\
 50718.7031 &   0.214 & 50718.7046 &   0.206 & 50721.6659 &   0.199 & 50721.6674 &   0.205 & 50721.6689 &   0.205 & 50725.6388 &   0.219 \\
 50725.6402 &   0.232 & 50725.6417 &   0.224 & 50726.6663 &   0.181 & 50726.6677 &   0.188 & 50726.6693 &   0.185 & 50727.6721 &   0.202 \\
 50727.6735 &   0.209 & 50727.6751 &   0.196 & 50728.6343 &   0.215 & 50728.6358 &   0.215 & 50728.6373 &   0.213 & 50730.6644 &   0.210 \\
 50730.6659 &   0.212 & 50730.6674 &   0.209 & 50752.5823 &   0.200 & 50752.5838 &   0.194 & 50753.5974 &   0.207 & 50753.5989 &   0.220 \\
 50753.6004 &   0.214 & 50754.5968 &   0.294 & 50754.5998 &   0.293 & 50755.5960 &   0.213 & 50755.5974 &   0.219 & 50755.5990 &   0.216 \\
 50756.5954 &   0.211 & 50756.5968 &   0.221 & 50756.5984 &   0.217 &    \nodata & \nodata &    \nodata & \nodata &    \nodata & \nodata \\
\enddata
\end{deluxetable}

\begin{deluxetable}{ccc}
\tablewidth{0pt}
\tablecaption{Photometric properties and effective temperature determination of V380 Cyg \label{tab:teff}}
\tablehead{
\colhead{Johnson\tablenotemark{a}}&
\colhead{Str\"omgren\tablenotemark{b}}&
\colhead{Vilnius\tablenotemark{c}}}
\startdata
$V=5.68\pm0.02$     &  $b-y=0.031$   & $U-P=0.20$                    \\
$B-V=-0.06\pm0.01$  &  $m_1=0.039$   & $P-Y=0.37$                    \\
$U-B=-0.76\pm0.02$  &  $c_1=0.111$   & $Y-V=0.22$                    \\
                    &  $\beta=2.587$ &                               \\
\tableline
$E(B-V)=0.18$       &  $E(B-V)=0.19$ & $E(B-V)=0.23$                 \\
$T_{\rm eff}=22\,700$ K (F96) &  $T_{\rm eff}=20\,200$ K & $T_{\rm eff}=20\,900$ K   \\
$T_{\rm eff}=24\,600$ K (P80) & & \\
$T_{\rm eff}=23\,100$ K (B81) & & \\
\enddata
\tablenotetext{a}{Photometry from this work. $E(B-V)$ computed using the 
intrinsic $(U-B)$ {\em vs} $(B-V)$ relation from Schmidt-Kaler 1982 
(luminosity class III) and the standard reddening slope of $E(U-B)/E(B-V)=
0.72$. Effective temperatures obtained from $T_{\rm eff}$ vs. $(B-V)_{\circ}$ 
calibrations of Flower 1996 (F96), Popper 1980 (P80) and B\"ohm-Vitense 
1981 (B81).}
\tablenotetext{b}{Photometry from Hauck \& Mermilliod 1998. $E(B-V)$ determined
from Crawford 1978 --- as explained in Figueras, Torra, \& Jordi 1991 and 
Jordi et al. 1997 --- and temperature computed by means of the photometric 
grids of R. Napiwotzki 1998 (private communication), based on Kurucz {\em 
ATLAS9} atmosphere models. Coefficients for the color excesses taken from
Strai\v{z}ys 1992 and Crawford \& Mandwewala 1976.}
\tablenotetext{c}{Photometry from S\={u}d\v{z}ius et al. 1992. Calibrations
in Strai\v{z}ys (1992) used for the calculation of the color excess and the 
effective temperature.}
\end{deluxetable}

\begin{deluxetable}{lr}
\tablewidth{0pt}
\tablecaption{Results from the light curve, radial velocity 
curve and UV/optical spectrophotometry analyses. 
\label{tab:prop}}
\tablehead{\colhead{Parameter} & \colhead{Value}}
\startdata 
\multicolumn{2}{c}{Light curve} \\
\tableline
Period (days)                                        &$12.425719\pm0.000014$ \\
Eccentricity                                         & $0.234\pm0.006$       \\
Inclination (deg)                                    & $82.4\pm0.02$         \\
$\omega$ (1989.0) (deg)                              & $132.7\pm0.3$         \\
$\dot{\omega}$ (deg/100 yr)                          & $24.0\pm1.8$          \\
$\frac{{T_{\rm eff}}_{\rm S}}{{T_{\rm eff}}_{\rm P}}$& $0.96\pm0.02$         \\
$ \left. \frac{L_{\rm P}}{L_{\rm S}} \right|_{U}$    & $19.14\pm0.11$        \\
$ \left. \frac{L_{\rm P}}{L_{\rm S}} \right|_{B}$    & $17.48\pm0.08$        \\
$ \left. \frac{L_{\rm P}}{L_{\rm S}} \right|_{V}$    & $17.31\pm0.07$        \\
$r_{\rm P}$                                          & $0.2478\pm0.0014$     \\
$r_{\rm S}$                                          & $0.0631\pm0.0008$     \\
\tableline
\multicolumn{2}{c}{Radial velocity curve\tablenotemark{a}} \\
\tableline
$K_{\rm P}$ (km~s$^{-1}$)                            & $94.5\pm1.5$          \\
$K_{\rm S}$ (km~s$^{-1}$)                            & $151.1\pm3.0$         \\
$q\equiv\frac{M_{\rm S}}{M_{\rm P}}$                 & $0.626\pm0.041$       \\
$\gamma$ (km~s$^{-1}$)                               & $2.4\pm1.0$           \\
$a$ (R$_{\odot}$)                                    & $59.2\pm0.8$          \\
\tableline
\multicolumn{2}{c}{UV/Optical Spectrophotometry} \\
\tableline
${T_{\rm eff}}_{\rm P}$ (K)                          & $21\,350\pm400$       \\
${v_{\rm micro}}_{\rm P}$ (km s$^{-1}$)              & $12\pm1$              \\
${v_{\rm micro}}_{\rm S}$ (km s$^{-1}$)              & $0$                   \\
$[m/H]$ (dex)                                        & $-0.44\pm0.07$        \\
$E(B-V)$ (mag)                                       & $0.17\pm0.02$         \\
$\log (R_{\rm P}^2/d^2)$                             & $-18.961\pm0.031$     \\
\enddata
\tablenotetext{a}{From Paper~I.}
\end{deluxetable}

\begin{deluxetable}{lrr}
\tablewidth{0pt}
\tablecaption{Properties of the V380~Cyg components. \label{tab:com}}
\tablehead{\colhead{}&\colhead{Primary}&\colhead{Secondary}}
\startdata
Spectral Type\tablenotemark{a}           & B1.5 II-III   & B2 V             \\
Mass (M$_{\odot}$)                       & $11.1\pm0.5$  & $6.95\pm0.25$    \\
Radius (R$_{\odot}$)                     & $14.7\pm0.2$  & $3.74\pm0.07$    \\
$\log g$ (cgs)                           & $3.148\pm0.023$&$4.133\pm0.023$  \\
$T_{\rm eff}$ (K)                        & $21\,350\pm400$&$20\,500\pm500$  \\
$\log (L/L_{\odot})$\tablenotemark{b}    & $4.60\pm0.03$  & $3.35\pm0.04$   \\
$M_{\rm bol}$\tablenotemark{c}~ (mag)    & $-6.75\pm0.07$ & $-3.62\pm0.10$  \\
$V$\tablenotemark{d}~ (mag)              & $5.74\pm0.02$  & $8.84\pm0.02$   \\
$d$\tablenotemark{e}~ (pc)               & \multicolumn{2}{c}{$1000\pm40$}  \\
$M_{\rm v}$\tablenotemark{f}~ (mag)      & $-4.79\pm0.10$ & $-1.69\pm0.10$  \\
$v_{\rm rot}$\tablenotemark{g}~ (km s$^{-1}$) & $98\pm4$ & $32\pm6$         \\
${v_{\rm rot}}_{\rm p-sync}$\tablenotemark{h}~ (km s$^{-1}$)&$99\pm2$&$25\pm2$\\
$\overline{\rho}$ (g cm$^{-3}$)          &$0.0049\pm0.0003$&$0.187\pm0.013$ \\
$\rho_{\rm c}$\tablenotemark{i}~ (g cm$^{-3}$)&$3.09\pm0.24$&$19.6\pm1.6$   \\
\enddata
\tablenotetext{a}{Hill \& Batten 1984.}
\tablenotetext{b}{From $L = 4 \pi R^2 \sigma T_{\rm eff}^4$.}
\tablenotetext{c}{From $\log (L/L_{\odot})$ and adopting 
${{\rm M}_{\rm bol}}_{\odot}=4.75$.}
\tablenotetext{d}{Computed from the out-of-eclipse magnitude and $(L_{\rm P}/
L_{\rm S})|_{\rm V}$.}
\tablenotetext{e}{Using $\log (R_{\rm P}^2/d^2)$ from the UV/optical 
spectrophotometry fit and the observed $R_{\rm P}$.}
\tablenotetext{f}{From $M_{\rm v} = V - 3.1 E(B-V) + 5 - 5 \log d$. Note that 
our values are completely consistent with what is expected for the spectral 
classification of the components. In addition, $M_{\rm bol}-M_{\rm v}$ for 
both components is in excellent agreement with the bolometric correction
calibration of Flower (1996).}
\tablenotetext{g}{Computed from Lyubimkov et al. 1996.}
\tablenotetext{h}{Pseudo-synchronization rotational velocity.}
\tablenotetext{i}{Following Kopal 1959 and using the internal concentration
parameters $k_2$.}
\end{deluxetable}

\begin{deluxetable}{lcl}
\tablewidth{0pt}
\tablecaption{Summary of time of minima determination for V380 Cyg. 
\label{tab:ecl}}

\tablehead{\colhead{HJD}&\colhead{Type\tablenotemark{a}}&\colhead{Reference}}
\startdata
$2423587.184\pm0.009$ & P & Kron 1935 \\
$2423592.565\pm0.007$ & S & Kron 1935 \\
$2438274.372\pm0.003$ & P & Semeniuk 1968 \\
$2438279.571\pm0.020$ & S & Semeniuk 1968 \\
$2441256.541\pm0.003$ & P & Battistini et al. 1974 \\
$2441261.620\pm0.004$ & S & Battistini et al. 1974 \\
$2443791.412\pm0.010$ & P & Dorren \& Guinan (unpub.) \\
$2443796.416\pm0.010$ & S & Dorren \& Guinan (unpub.) \\
$2447543.936\pm0.005$ & P & This study (APT-Phoenix10) \\
$2447548.931\pm0.005$ & S & This study (APT-Phoenix10) \\
$2450029.125\pm0.005$ & P & This study (FCC 0.8-m APT) \\
$2450034.033\pm0.005$ & S & This study (FCC 0.8-m APT) \\
\enddata
\tablenotetext{a}{P and S stand for primary and secondary minimum, 
respectively.}
\end{deluxetable}

\begin{deluxetable}{ccccccccc}
\tablewidth{0pt}
\tablecaption{Model predictions with different amounts of overshooting for 
the components of V380~Cyg. \label{tab:mod}}
\tablehead{\colhead{}&\colhead{}&\multicolumn{3}{c}{Primary comp.}&\colhead{}&
\multicolumn{3}{c}{Secondary comp.}}
\startdata
$\alpha_{\rm ov}$&&Age (Myr)& $\log T_{\rm eff}$ & $\log k_2$ 
&& Age (Myr) & $\log T_{\rm eff}$ & $\log k_2$ \\
\tableline
$0.2$&&$20.5\pm1.5$&$4.267\pm0.010$&$-2.66\pm0.03$&&$20\pm3$&$4.311\pm0.009$&$-2.10\pm0.02$\\
$0.4$&&$22.9\pm1.5$&$4.282\pm0.010$&$-2.75\pm0.03$&&$23\pm3$&$4.313\pm0.009$&$-2.11\pm0.02$\\
$0.6$&&$25.2\pm1.5$&$4.324\pm0.010$&$-2.91\pm0.03$&&$26\pm3$&$4.314\pm0.009$&$-2.11\pm0.02$\\
\enddata
\end{deluxetable}


\begin{thebibliography}{}
\bibitem[Alcock et al. (1997)]{Aea97}
Alcock, C., Allsman, R. A., Alves, D. et al. 1997, AJ, 114, 326
\bibitem[Andersen (1991)]{A91}
Andersen, J. 1991, A\&AR, 3, 91
\bibitem[Andersen et al. (1990)]{Aea90}
Andersen, J., Nordstr\"om, B., \& Clausen, J. V. 1990, ApJ, 363, L33
\bibitem[Batten (1962)]{B62}
Batten, A. H. 1962, Pub. Dom. Astropys. Obs. 12, 91
\bibitem[Battistini et al. (1974)]{Bea74}
Battistini, P., Bonifazi, A., \& Guarnieri, A. 1974, Ap\&SS, 30, 163
\bibitem[Bohlin (1996)]{B96}
Bohlin, R. 1996, AJ, 111, 1743
\bibitem[B\"ohm-Vitense (1981)]{BV81}
B\"ohm-Vitense, E. 1981, Ann. Rev. A\&A, 19, 295 (B81)
\bibitem[Canuto (1999)]{C99}
Canuto, V. M. 1999, ApJ, 518, L119
\bibitem[Canuto \& Mazzitelli (1991)]{CM91}
Canuto, V. M., \& Mazzitelli, I. 1991, ApJ, 370, 295
\bibitem[Canuto \& Mazzitelli (1992)]{CM92}
Canuto, V. M., \& Mazzitelli, I. 1992, ApJ, 389, 724
\bibitem[Claret (1995)]{C95}
Claret, A. 1995, A\&AS, 109, 441
\bibitem[Claret (1999)]{Cl99}
Claret, A. 1999, A\&A, 350, 56
\bibitem[Claret \& Gim\'enez (1992)]{CG92}
Claret, A., \& Gim\'enez, A. 1992, A\&AS, 96, 255
\bibitem[Claret \& Gim\'enez (1993)]{CG93}
Claret, A., \& Gim\'enez, A. 1993, A\&A, 277, 487
\bibitem[Claret et al. (1995)]{Cea95}
Claret, A., Gim\'enez, A., \& Cunha, N. C. S. 1995, A\&A, 299, 724
\bibitem[Crawford (1978)]{C78}
Crawford, D. L. 1978, AJ, 83, 48
\bibitem[Crawford \& Mandwewala (1976)]{CM76}
Crawford, D. L., \& Mandwewala, N. 1976, PASP, 88, 917
\bibitem[Deupree (1998)]{D98}
Deupree, R. G. 1998, ApJ, 499, 340
\bibitem[Edvardsson et al. (1993)]{Eea93}
Edvardsson, B., Andersen, J., Gustafsson, B., Lambert, D. L., Nissen, P. E.,
\& Tomkin, J. 1993, A\&A, 275, 101
\bibitem[ESA (1997)]{E97}
ESA 1997, The Hipparcos and Tycho Catalogues, ESA SP-1200
\bibitem[Figueras et al. (1991)]{Fea91}
Figueras, F., Torra, J., \& Jordi, C. 1991, A\&AS, 87, 319
\bibitem[Fitzpatrick 1999]{F99}
Fitzpatrick, E. L. 1999, PASP, 111, 63
\bibitem[Fitzpatrick \& Massa 1990]{FM90}
Fitzpatrick, E. L., \& Massa, D. 1990, ApJS, 72, 163
\bibitem[Fitzpatrick \& Massa (1999)]{FM99}
Fitzpatrick, E. L., \& Massa, D. 1999, ApJ, 525, 1011 (FM99)
\bibitem[Flower (1996)]{F96}
Flower, P. J. 1996, ApJ, 469, 355 (F96)
\bibitem[Gim\'enez (1984)]{G84}
Gim\'enez, A. 1984, in Observational tests of the Stellar Evolution Theory,  
eds. A. Maeder \& A. Renzini (Dordrecht: Reidel), 419
\bibitem[Gim\'enez (1985)]{G85}
Gim\'enez, A. 1985, ApJ, 297, 405
\bibitem[Gim\'enez \& Bastero (1995)]{GB95}
Gim\'enez, A., \& Bastero, M. 1995, Ap\&SS, 226, 99
\bibitem[Gim\'enez \& Garc\'{\i}a-Pelayo (1983)]{GGP83}
Gim\'enez, A., \& Garc\'{\i}a-Pelayo, J. M. 1983, Ap\&SS, 92, 203
\bibitem[Gim\'enez et al. (1994)]{Gea94}
Gim\'enez, A., Claret, A., \& Guinan, E. F. 1994, presented in the 22nd
General Assembly of the IAU (JD 12), The Hague.
\bibitem[Gim\'enez et al. (1999)]{Gea99}
Gim\'enez, A., Guinan, E. F., \& Montesinos, B. 1999, Stellar Structure: 
Theory and Test of Connective Energy Transport, ASP Conf. Ser., Vol. 173
(San Francisco: ASP)
\bibitem[Grison et al. (1995)]{Gea95}
Grison, P., Beaulieu, J. -P., Pritchard, J. D. et al. 1995, A\&AS, 109, 447 
\bibitem[Guinan \& Maloney (1985)]{GM85}
Guinan, E. F., \& Maloney, F. P. 1985, AJ, 90, 1519
\bibitem[Guinan et al. (1998)]{Gea98}
Guinan, E. F., Fitzpatrick, E. L., DeWarf, L. E., Maloney, F. P., Maurone,
P. A., Ribas, I., Gim\'enez, A., Pritchard, J. D., \& Bradstreet, D. H. 1998,
ApJ, 509, L21
\bibitem[Hauck \& Mermilliod (1998)]{HM98}
Hauck, B., \& Mermilliod, M. 1998, A\&AS, 129, 431
\bibitem[Hill \& Batten (1984)]{HB84}
Hill, G., \& Batten, A.H. 1984, A\&A, 141, 39
\bibitem[Jordi et al. (1997)]{Jea97}
Jordi, C., Masana. E., Figueras, F., \& Torra, J. 1997, A\&AS, 123, 83
\bibitem[Kaluzny et al. (1998)]{Kea98}
Kaluzny, J., Stanek, K. Z., Krockenberger, M., Sasselov, D. D., Tonry, J. L.,
\& Mateo, M. 1998, AJ, 115, 1016
\bibitem[Kaluzny et al. (1999)]{Kea99}
Kaluzny, J., Mochejska, B. J., Stanek, K.Z., Krockenberger, M., Sasselov, D. D.,
Tonry, J. L., \& Mateo, M. 1999, AJ, 118, 346
\bibitem[Kilian et al. (1994)]{Kea94}
Kilian, J., Montenbruck, O., \& Nissen, P. E. 1994, A\&A, 284, 437
\bibitem[Kopal (1959)]{K59}
Kopal, Z. 1959, Close binary systems (New York: Wiley)
\bibitem[Kron (1935)]{K35}
Kron, G. E. 1935, ApJ, 82, 225
\bibitem[Kurucz (1991)]{K91} 
Kurucz, R. L. 1991, in Stellar Atmospheres: Beyond Classical Models, eds. 
L. Crivellari et al. (Dordrecht: Kluwer), 441
\bibitem[Kurucz (1994)]{K94}
Kurucz, R. L. 1994, CD-ROM No 19
\bibitem[Lyubimkov et al. (1996)]{Lea96}
Lyubimkov, L. S., Rachkovskaya, T. M., Rostopchin, S. I., \& Tarasov, A. E.
1996, Astron. Rep., 40, 46
\bibitem[Massa \& Fitzpatrick (2000)]{MF00}
Massa, D., \& Fitzpatrick, E. L. 2000, ApJS, 126, 517
\bibitem[Meyer et al. (1998)]{Mea98}
Meyer, D. M., Jura, M., \& Cardelli, J. A. 1998, ApJ, 493, 222
\bibitem[Milone et al. (1992)]{Mea92}
Milone, E. F., Stagg, C. R., \& Kurucz, R. L. 1992, ApJS, 79, 123
\bibitem[Milone et al. (1994)]{Mea94}
Milone, E. F., Stagg, C. R., Kallrath, J., \& Kurucz, R. L. 1994, BAAS, 184, 
0605
\bibitem[Nichols \& Linksy (1996)]{NL96}
Nichols, J. S., \& Linksy, J. L. 1996, AJ, 111, 517
\bibitem[Pols et al. (1997)]{Pea97}
Pols, O. R., Tout, C. A., Schr\"oder, K. -P., Eggleton, P. P., \& Manners, J.
1997, MNRAS, 289, 869
\bibitem[Popper (1980)]{P80}
Popper, D. M. 1980, Ann. Rev. A\&A, 18, 115 (P80)
\bibitem[Popper (1981)]{P81}
Popper, D. M. 1981, ApJS, 47, 339
\bibitem[Popper \& Guinan (1998)]{PG98}
Popper, D. M., \& Guinan, E. F. 1998, PASP, 110, 572 (Paper~I)
\bibitem[Ribas et al. (1999)]{Rea99}
Ribas, I., Gim\'enez, A., Jordi, C., Claret, A., \& Guinan, E. F. 1999,
in ASP Conf. Ser. 173, Stellar Structure: Theory and Test of Connective 
Energy Transport, eds. A. Gim\'enez, E. F. Guinan, \& B. Montesinos,
(San Francisco: ASP), 253
\bibitem[Ribas et al. (2000a)]{Rea00a}
Ribas, I., Jordi, C., Torra, J., \& Gim\'enez, A. 2000a, MNRAS, 313, 99
\bibitem[Ribas et al. (2000b)]{Rea00b}
Ribas, I., Jordi, C., \& Gim\'enez, A. 2000b, MNRAS, submitted
\bibitem[Semeniuk (1968)]{S68}
Semeniuk, I. 1968, AcA, 18, 1
\bibitem[Schmidt-Kaler (1982)]{SK82}
Schmidt-Kaler, T. 1982, Landolt-B\"ornstein, Vol. II, 453
\bibitem[Schr\"oder et al. (1997)]{Sea97}
Schr\"oder, K. -P., Pols, O. R., \& Eggleton, P. P. 1997, MNRAS, 285, 696
\bibitem[Stanek et al. (1998)]{Sea98}
Stanek, K. Z., Kaluzny, J., Krockenberger, M., Sasselov, D. D., Tonry, J. L.,
\& Mateo, M. 1998, AJ, 115, 1894
\bibitem[Stanek et al. (1999)]{Sea99}
Stanek, K. Z., Kaluzny, J., Krockenberger, M., Sasselov, D. D., Tonry, J. L.,
\& Mateo, M. 1999, AJ, 117, 2810
\bibitem[Stothers (1974)]{S74}
Stothers, R. 1974, ApJ, 194, 651
\bibitem[Strai\v{z}ys (1982)]{S82}
Strai\v{z}ys, V. 1992, Multicolor Stellar Photometry (Tucson: Pachart Pub.
House)
\bibitem[S\={u}d\v{z}ius \& Bobinas (1992)]{SB92}
S\={u}d\v{z}ius, J., \& Bobinas, V. 1992, Bull. Vilnius Obs., 86, 59
\bibitem[Tassoul (1987)]{T87}
Tassoul, J. -L. 1987, ApJ, 322, 856
\bibitem[Tassoul (1988)]{T88}
Tassoul, J. -L. 1988, ApJ, 324, L71
\bibitem[Udalski et al. (1998)]{Uea98}
Udalski, A., Soszy\'nski, I., Szyma\'nski, M., Kubiak, M., Pietrzy\'nski, G.,
Wo\'zniak, P., \& \.{Z}ebru\'n, K. 1999, AcA, 48, 563
\bibitem[Wilson \& Devinney (1971)]{WD71}
Wilson, R. E., \& Devinney, E. J. 1971, ApJ, 166, 605 (WD)
\end{thebibliography}
\end{document}